\documentclass{aa}
\usepackage{natbib}
\bibpunct{(}{)}{;}{a}{}{,} 
\usepackage{adjustbox}
\usepackage{graphicx}
\usepackage{txfonts}
\usepackage{ragged2e}

\usepackage[colorlinks=true,     linkcolor=blue, citecolor=blue, filecolor=blue, urlcolor=blue]{hyperref}

\begin{document}



\title
    {
    The YMDB catalog: Young massive detached binaries for the determination of high-precision absolute stellar parameters\thanks{Full version of Tables 2, 3 and 4 are only available in electronic form at the CDS via anonymous ftp to cdsarc.u-strasbg.fr (130.79.128.5) or via http://cdsweb.u-strasbg.fr/cgi-bin/qcat?J/A+A/.}
    }
\titlerunning{The YMDB catalog}


\author{
    Pablo Mart\'in-Ravelo 
    \inst{1,2}
    \and
    Roberto Gamen 
    \inst{3,4}
    \and 
    Julia I. Arias 
    \inst{1}
    \and 
    Andr\'e-Nicolas Chen\'e 
    \inst{2}
    \and 
    Rodolfo H. Barb\'a \thanks{In Memoriam (1962--2021)}
    \inst{1}
    }
\institute{Departamento de Astronom\'{\i}a, Universidad de La Serena, Av. Cisternas 1200 Norte, La Serena, Chile.\\
    \and 
    Gemini Observatory/NSF’s NOIRLab, 670 N. A‘ohoku Place, Hilo, HI 96720, USA \\
    \email{pablo.martin@noirlab.edu}
    \and 
    Instituto de Astrof\'{\i}sica de La Plata, CONICET--UNLP, Paseo del Bosque s/n, 1900, La Plata, Argentina.\\
    \and 
    Facultad de Ciencias Astron\'omicas y Geof\'{\i}sicas, Universidad Nacional de La Plata, Argentina.
    }

\date{Received: 20 June 2024 / Accepted: 05 July 2024}

\abstract
    {Massive stars play a crucial role in the cosmic dynamics and chemical evolution of galaxies. Despite their significance, our understanding of their evolution and properties remains limited. An accurate determination of stellar parameters, such as the mass and radius, is essential for advancing our knowledge. Detached eclipsing binaries (DEBs) are particularly valuable for these determinations due to the minimal interaction between their stellar components, allowing for precise measurements.}
    {This study aims to introduce the Young Massive Detached Binary (YMDB) catalog, designed to address the gap in the high-precision absolute parameter determination for young massive stars. By focusing on DEBs within the spectral range O9-B1, this catalog seeks to provide a reliable database for future astronomical studies and improve our understanding of massive star evolution.}
    {We conducted a photometric analysis of 87 young massive stars in detached eclipsing systems using TESS light curves (LCs) that were processed through a custom pipeline. This analysis involved determining the amplitude of magnitude variations, orbital periods, times of minima, eccentricities, and the presence of apsidal motion and heartbeat phenomena. A thorough literature review was performed to obtain MK spectral classifications. We performed our own spectral classification of 19 systems to support the sample where a new classification was lacking or inconclusive.}
    {The analysis identified 20 previously unreported binary systems, with 13 newly recognized as variable stars. Among the 87 stars examined, 30 are confirmed as YMDB members, and 25 are candidates pending spectral classification. The exclusion of the remaining 32 stars is attributed to unsuitable spectral types or their nondetached binary nature. Notable findings include the identification of new LC classifications, eccentricities in 13 systems, and heartbeat phenomena in several targets.}
    {The YMDB catalog offers a resource of high-quality LCs and reliable stellar classifications, serving as a valuable tool for the astronomical community.}

    
\maketitle

\section{Introduction}
Massive stars, defined by their cataclysmic end as core collapse supernovas, are typically those with an initial mass of eight or more solar masses and they fall into the OB spectral classification. These stellar behemoths play a crucial role in cosmic dynamics and chemical evolution, with their supernova events significantly enriching the interstellar medium with heavy elements. Understanding massive stars is fundamental to comprehending stellar evolution, the evolutionary history of galaxies, and the Universe at large. Due to their immense luminosity, they are key observable objects in distant galaxies, making them essential for astronomical studies with both current and forthcoming space and ground-based large telescopes.

Our current understanding of massive star evolution remains limited, as evidenced by the persistent mass discrepancy between empirical measurements from orbital dynamics and theoretical model predictions. This discrepancy has been a longstanding issue in astrophysics \citep[see e.g.,][and references therein]{1992A&A...261..209H,2010A&A...524A..98W}. Factors inherent in binary systems, such as stellar rotation, which influence orbital ellipticity, synchronization, tidal effects, limb darkening, and radiative propagation, are crucial in this context. Evolutionary models that incorporate rotation show marked differences compared to nonrotational models. Additionally, research by \citet{2013A&A...560A..16M} highlights significant variations in how current models handle mass loss in these stars.

The determination of reliable masses for massive stars is only achievable in binary systems, particularly through the study of light curves (LCs) and radial velocity (RV) curves of eclipsing binaries. The likelihood of encountering massive stars within binary systems is fairly high, especially during their main sequence phase, the most extended period of a star's lifecycle. Comprehensive spectroscopic surveys reveal that around 75\% of main sequence O-type stars \citep[cf.][]{2017IAUS..329...89B}, and a similar ratio of early B-type stars \citep{2012MNRAS.424.1925C}, are part of gravitationally bound systems.

The endemic multiplicity among massive stars offers unique research opportunities, although the method has its limitations. An inherent consequence of binarity is the interaction among its components, which make their evolution paths deviate from those of solitary stars of similar types. This interaction issue was emphasized by \citet{2012Sci...337..444S}, who found that at least 71\% of O-type stars in binary systems interact with their companions during their lifetimes, with 29\% eventually merging into a single entity. Detecting past interactions poses challenges, as evidenced by discrepancies between empirical mass measurements and predictions from atmospheric or stellar evolution models, which often overlook these interactions. 

Detached eclipsing binaries (DEBs) stand as critical objects in this context. These systems, charaterized by a negligible or minimal interaction between their stellar components, perfectly meet the criteria for accurate stellar analysis. Given that most massive stars are part of multiple systems, it is preferable to focus on those with well-documented multiplicity. The best single stars for study are often the components of binary systems, as proposed by \cite{2011BSRSL..80..543D} and further discussed by \cite{2021FrASS...8...53E}. DEBs provide a unique opportunity to determine the physical properties of stars with a high accuracy, such as their masses, and radii. Moreover, they can be used to determine distances. In this context, DEBs are key objects. Determinations from DEBs form the basis for calibrating crucial relationships among stellar parameters, such as the mass--luminosity relationship.

Among massive stars, those within the O8--B3 spectral-type range are predominant (just as a consequence of the stellar mass distribution). However, only a limited number have had their absolute parameters accurately determined. 
Our research aims to address this gap by examining a selection of DEBs within this spectral range. We intend to analyze their LCs and RV curves to derive precise stellar parameters. 

In this work, we present a curated database of young massive stars, with spectral types around B0~V, in DEBs. This endeavor is the first step to maintain an up-to-date database of empirical absolute parameters for massive stars. To achieve our objectives, we have carefully selected systems from a careful search in the available literature (see details in Sect.~\ref{sample}). After that, we generated the LCs of each target in the database using data from the NASA Transiting Exoplanet Survey Satellite (TESS), and analyzed their variations to identify DEBs. This is explained in Sect.~\ref{tess}. TESS LCs were also examined for periodicities, to determine eccentricities, ephemerides, and also for the detection of pulsation and/or heartbeat phenomena. Some targets were spectroscopically observed to obtain their spectral classification, which were confusing in the literature. Observations are described in Sect.~\ref{casleo}. All of the results that populate the Young Massive Detached Binaries (YMDB) catalog are shown in Sect.~\ref{results}.

\section{General methodology for the YMDB catalog}
Our study's approach integrates a thorough search and analysis methodology to identify and evaluate candidates for detailed investigation. Starting with an extensive review of databases and literature, we refined our list of potential candidates by examining their spectral classifications and the likelihood of misclassification, and by quickly reviewing raw TESS LCs to exclude any undoubted nondetached binaries. This process narrowed our initial pool to 87 candidates.

In preparing the LCs, we initially set up background and target masks using a structured pixel grid to minimize light contamination, which is vital for precise LC extraction. We continuously evaluated data quality and implemented polynomial fitting to correct known and recurrent flux pattern variations in the TESS data, such as rollovers, ensuring a cleaner and more accurate representation of the LCs. 

We utilized Gaussian Fitting (GF) and Box Least Square (BLS) methods to accurately determine orbital periods and other temporal features. Further analysis of the LCs involved meticulous visual inspection to confirm the detached nature of systems and identify any additional features such as pulsations or heartbeats.

To enhance the validity of the spectral classifications, new spectroscopic observations were conducted using the Jorge Sahade telescope at CASLEO. These efforts were focused on acquiring low-resolution spectra, which were processed using established methods, to clarify any existing classification uncertainties.

\subsection{Candidates selection} \label{sample}
Candidates for the study were identified through extensive searches in various databases and literature. These sources included the Spectroscopic Binary Orbits Ninth Catalog \citep{2004A&A...424..727P}, OWN Survey \citep{2007BAAA...50..105G,2008RMxAC..33...54G,2017IAUS..329...89B}, IACOB \citep{2011BSRSL..80..514S}, Eclipsing Variables Catalog \citep{2013AN....334..860A}, and the multiplicity of northern O-type spectroscopic systems project \citep[MONOS;][]{2019AnA...626A..20M,2021A&A...655A...4T}, among others. Additionally, potential targets suggested by collaborative efforts within the astronomical community were also considered.

The search criteria encompassed a wide range, including systems with components in the O7--B3 III--V spectral class, to account for potential misclassifications. This cautious approach was used to ensure all potentially relevant systems were considered, particularly those that might align with the desired O9-B1 IV-V classification upon closer scrutiny. 
A thorough review of the bibliography was conducted to assess the reliability of the spectral classifications, the availability of LCs, and the reported detachment status in the literature, narrowing down the original list of 339 targets to 186 candidates.

The process continued with an analysis of TESS data to eliminate any candidates showing clear non-Algol type variability (i.e, Beta Lyr\ae\, and W Urs\ae\, Majoris) reducing the number to 114 candidates. A custom pipeline was then developed for extracting high quality TESS LCs, allowing for the estimation of orbital periods (P) and minima (T0), as well as the identification of eccentric orbits, heartbeats and apsidal motion throughout the TESS sectors. This was achieved using both GF and BLS methods. The candidates were meticulously scrutinized, focusing on the identification of Algol-type variables. Systems not fitting the Algol type criteria were excluded, resulting in 87 stars qualifying as potential candidates which we present in this work.
Details of this task are given in the following.

\subsection{Construction of light curves} \label{tess}
In constructing the LC, the initial step involved assessing the quality masks for each TESS sector, related to the star, to determine whether the default rejection criteria were adequate. For the majority of LCs, it was found that the default quality mask met the needs effectively. 

\subsubsection{Background mask}
To construct a sky background mask for each sector, a 30$\times$30 pixel area centered on the target system is analyzed. Stars within this area, along with their magnitudes, are identified using Gaia DR3 data. The \textsc{lightkurve} tool \citep{2018ascl.soft12013L} is then employed to generate a mask with threshold zero, specifically excluding pixels affected by stars brighter than a determined magnitude limit. This upper limit is set at 5 magnitudes fainter than the brightest star within a 3-pixel radius of the target system, effectively minimizing contamination from other stars in the extraction of the desired LC.

\subsubsection{Target mask}
Following the construction of the sky background mask, a target mask is developed using a similar methodology. The function that delineates boxes around mapped stars based on their brightness is also applied to isolate the Target Pixel File (TPF) surrounding the star of interest. Adjustments to the size and position of this cut, alongside a specified threshold value, are made using the \textsc{lightkurve} tool to craft the target mask. This stage might require iteration, involving comparisons of the resulting LC with published ones and the LCs of individual pixels, to verify the accuracy of the selected parameters and the suitability of the target mask.

\subsubsection{Data selection}
The analysis of the unfolded extracted LC involves identifying and flagging problematic data that could compromise the construction of the LC. A visual inspection of the cadence for each TESS sector is essential, as it can reveal areas potentially affected by gaps in data—either due to TESS CCD readouts or data omitted by the quality mask. Data adjacent to these gaps might exhibit a different background profile, making parts of the LC unreliable if the background extraction fails to account for this variance. Additionally, these segments may experience slight magnitude shifts ($\Delta$mag) and may require independent normalization, posing challenges for systems with long periods.

\subsubsection{Determination of ephemeris} \label{ephemeris}
In determining the ephemeris of our studied systems, we incorporate two principal methods: GF and BLS. GF is primarily utilized for its simplicity and effectiveness in identifying eclipse minima by fitting Gaussian profiles, a method similar to the Phase Dispersion Minimization (PDM) approach.

Phase dispersion minimization is traditionally used to detect stable periodic signals by minimizing scatter across phased data, making it ideal for datasets like those from TESS which may include gaps or nonsinusoidal variations. By adapting this methodology, our GF process not only identifies the minima but also estimates the period from these minima's separations, similarly to how PDM assesses periodicity by examining the variance across different data bins.

After initial minima identification, we fold the LC and apply GF iteratively across all sectors, optimizing the ephemeris precision in a manner analogous to refining PDM's phase coverage by adjusting bin overlaps or employing smoother functions in updated PDM versions.

Box least square is used alongside GF for period verification, with both methods critically reviewed against known periods from literature when available in order to cross-verify with established data the reliability of our adapted PDM-inspired techniques in analyzing TESS's time series data.

\subsubsection{Phase-synchronized polynomial fitting}
Photometric observations from the TESS mission, similar to those from its predecessor Kepler, are susceptible to instrumental systematic trends that can obscure or distort the intrinsic stellar signals. These systematics arise primarily from spacecraft jitter and other operational imperfections, which introduce noise and trends across different timescales into the captured LCs. 

To mitigate these effects, we employed a polynomial fitting method, analogous to the Pixel Level Decorrelation (PLD) method utilized in the EVEREST pipeline for K2 data. This method has been demonstrated to effectively remove correlated noise due to spacecraft motion by fitting and subtracting systematics directly from the pixel-level data \citep{2016AJ....152..100L}. 

While our approach uses a simpler polynomial fitting of lower orders, it shares the core principle of modeling the observed data to isolate and remove instrumental signatures. The primary distinction of our method, which we designate as "Phase-Synchronized Polynomial Fitting" (PSPF), lies in its utilization of the previously known orbital periods to synchronize (fold) phase points within a TESS sector. This synchronization enables the calculation of fits for data points that share the same phase, thus forming a more complex function that better adapts to the unique signal trends of TESS data. Therefore, PSPF is suited for systems with orbital periods shorter than the duration of a TESS sector. The method cannot be applied if the period exceeds the sector length (PLD should be used in those cases), unless the signal is expected to remain constant in different phases of the LC, such as outside the eclipses in highly detached binaries. It becomes particularly reliable when the orbital period is substantially shorter than the sector length, allowing for enhanced signal integrity post-correction through multiple phase folds. 

The correction process begins with the identification of phase points across the LC that are consistent over multiple orbits, meaning without flux variations other than those related to the eclipsing nature of the system or other synchronized pulsations (whose periods are multiple of the orbital period). This leverages the comparative stability of shorter-period systems which may ignore not-synchronized variations if enough points per phase are provided. Each phase point will provide then a value not only for that phase, but for every other phase point in the LC. 
A weighted mean is calculated for these values in each phase point to construct a robust fit model. This model is then used to adjust the LC, with the median of all polynomial corrections applied to derive the final corrected LC. This procedure was only implemented when a consistent pattern of systematic error was evident across the entire sector, ensuring that the corrections made were both meaningful and substantiated by the data.

For about 20\% of the candidates in our study that exhibited clear systematic trends, the PSPF was crucial for assembling a clean LC. A polynomial fit of order 1 was mainly used, but the selection of polynomial order was tailored to each target's specific noise characteristics, with higher orders only used seldom and for more complex systematic patterns.

Our PSPF corrections were validated against already corrected LCs, showing a significant reduction in noise and the preservation of intrinsic astrophysical signals and effectively detrending the TESS LCs, akin to the results reported by the EVEREST pipeline when applying PLD to K2 data \citep{2018AJ....156...99L}.

\subsection{Light-curves analysis}
\begin{figure*}
	\includegraphics[width=2\columnwidth]{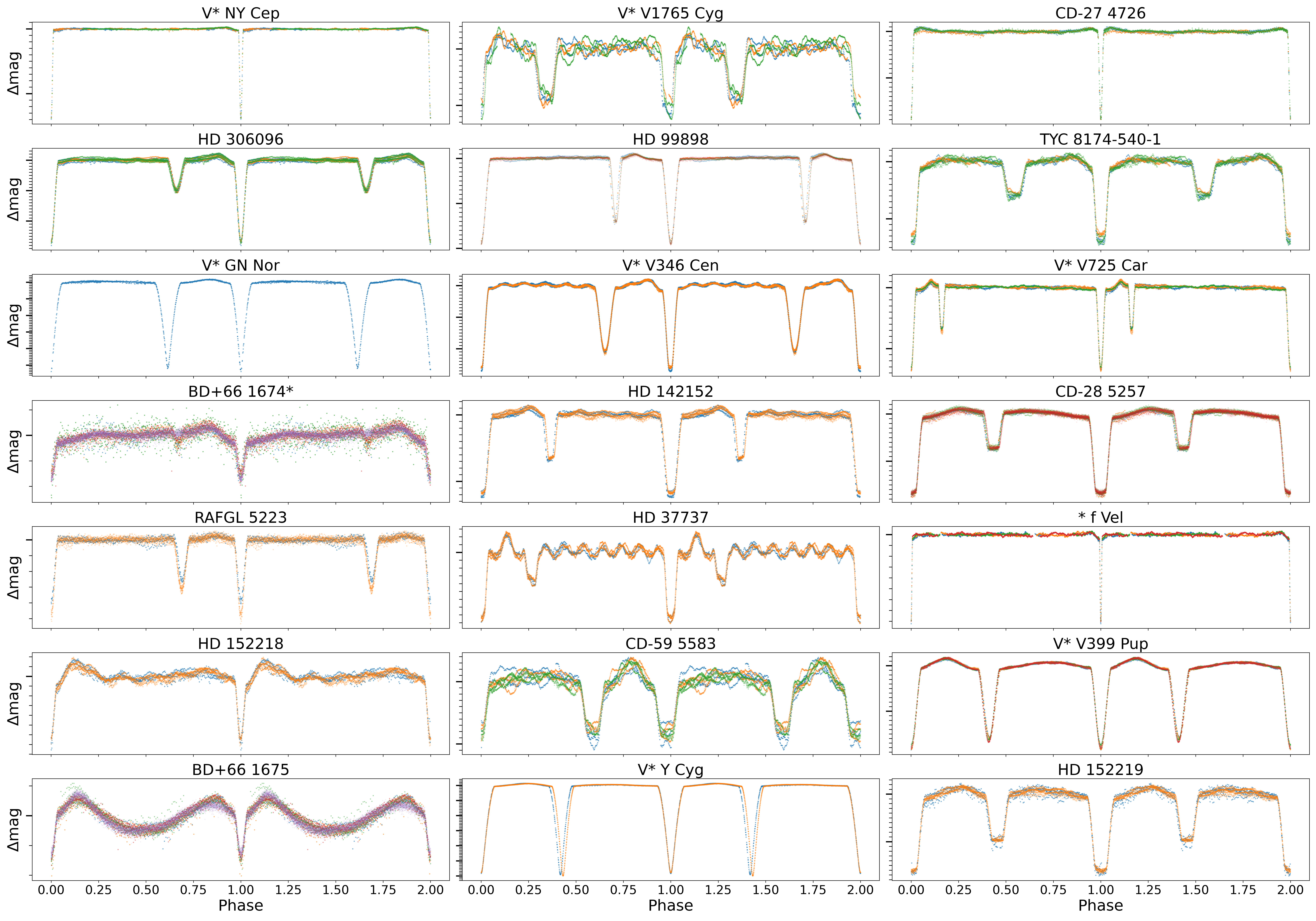} 
	\caption{Light curves of systems displaying heartbeat-like features. In each subpanel, bold, long ticks on the Y-axis denote increments of 0.1$\Delta mag$, and thin, short ticks indicate increments of 0.01$\Delta mag$.}
	\label{lc:Heartbeats}
\end{figure*}

All generated LCs underwent meticulous inspection to confirm their detached nature, identify any pulsations or heartbeats, and determine eccentricity and apsidal motion. Visual inspection was used to confirm the detached nature of the system, which was unequivocally determined by the distinct transitions marking the beginning and end of eclipses in the LC. Pulsational features, indicative of either intrinsic variability of the components or the presence of a complex multiple system, were also identified through visual inspection of the LCs. In cases of multiple systems, efforts were made to disentangle the LCs using templates crafted from median data selected, when possible, from segments of the LC minimally affected by eclipses or intrinsic pulsations of the main system. The heartbeat phenomenon was identified as a distinct photometric variation occurring during orbital phases with closer eclipses, indicative of the expected periastron. We compared those variations against all forms of heartbeat showed in \citet{2012ApJ...753...86T}. Detailed descriptions of these cases are provided in Section~\ref{heartbeat} and their LCs are shown in Figure \ref{lc:Heartbeats}.

For systems where eccentricity was not immediately obvious, we employed GF to precisely measure the centers of the primary and secondary eclipses, using a threshold of $\Delta\phi$=0.0002, to determine whether their phase separation differed from 0.5. It is important to note that an exact half-period interval between eclipses does not necessarily imply a circular orbit; the system could be eccentric with a periastron longitude of either 90$^\circ$ or 270$^\circ$. Therefore, values listed in Tables 2, 3, and 4 may imply eccentricity but never circularity. Variations in the fit to the secondary eclipse across different sectors were analyzed, and systems exhibiting changes beyond a certain threshold ($\Delta\phi$=0.0002) were flagged for potential apsidal motion. Additionally, since heartbeat phenomena occur exclusively in eccentric binaries, the detection of heartbeat signals in cases where only one eclipse was observed allowed us to identify the system as eccentric. 

\subsection{Spectroscopic observations and classification} \label{casleo}
To address the inconsistencies in spectral classifications found in the existing literature, we undertook the task of obtaining new spectra.
Low-resolution spectra were obtained using the Jorge Sahade telescope at Complejo Astronómico El Leoncito (CASLEO) in San Juan, Argentina, utilizing the REOSC spectrograph in single mode alongside the new Sophia CCD. This equipment yielded spectra with a resolution of $R\sim1000$, covering the wavelength range of 3760--5845~\AA\, suitable for spectral classification in the MKK system.
Spectral data processing was conducted in the standard way using \textsc{iraf}\footnote{NOIRLab IRAF is distributed by the Community Science and Data Center at NSF NOIRLab, which is managed by the Association of Universities for Research in Astronomy (AURA) under a cooperative agreement with the U.S. National Science Foundation.} routines.

\begin{figure*}[h!]
\centering
	\includegraphics[width=0.85\textwidth]{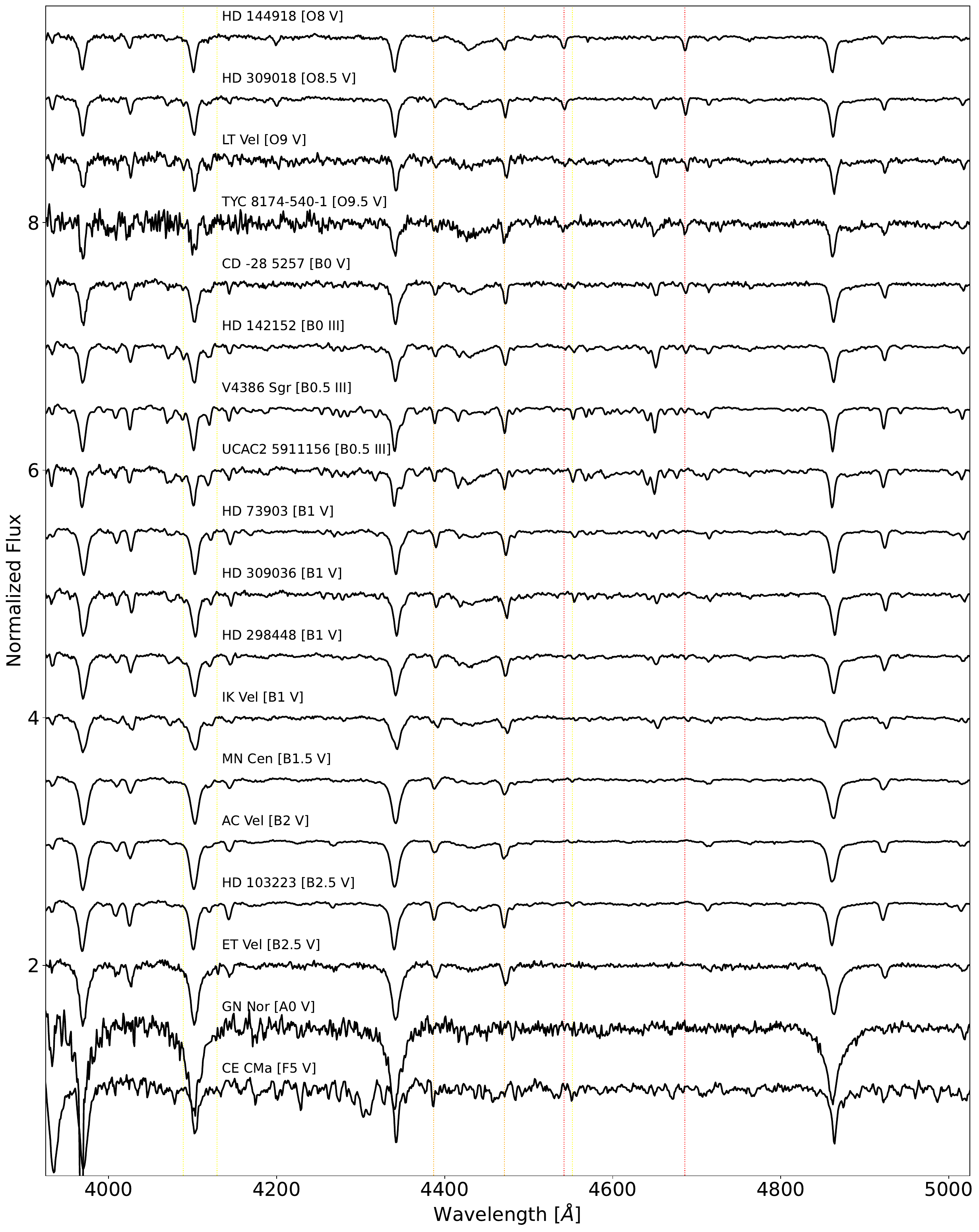} 
	\caption{CASLEO spectra classified in this work. Vertical lines indicate most important spectral lines used for classification, i.e., He \textsc{ii} (in red), He \textsc{i} (orange), and Si \textsc{ii--iii--iv} (yellow).}
	\label{fig:espectros}
\end{figure*}

The resulting spectra are depicted in Fig.~\ref{fig:espectros}. They have been classified following the criteria described in \citet{1990PASP..102..379W,2011ApJS..193...24S,2014ApJS..211...10S}. To summarize, we employed the ratio between He \textsc{ii} $\lambda$4541 and He \textsc{i} $\lambda$4387, for late O-types, and Si \textsc{iii} $\lambda$4552/Si \textsc{iv} $\lambda$4089, for early B-types. Further details about spectral classification of OB stars can be found in the references provided.

The new spectral types are presented in the corresponding Tables~\ref{tab:YMDB_confirmed},~\ref{tab:YMDB_candidates} and~\ref{tab:YMDB_unqualified}, column ST1, and are indicated with the label ``tw''. No ST2 is given due to the low spectral resolution, which prevents the disentangling of both components in the system.

\section{Results} \label{results}
Among the 87 systems, this study reveals 20 new eclipsing binaries (Table~\ref{tab:starlist_v2}), including 13 whose variable nature was previously unknown, five targets which were known as photometric variables or spectroscopic binaries, but not as DEB, and two of them were identified as nonthermal contact (because they present different light minima). Moreover, we introduce new LC classifications for 30 systems and report novel findings such as eccentricity in 13 systems, heartbeat features in 17, new types of variability in 11, and debut TESS LC presentations for the majority of the sample. 

The study's results are presented in three distinct tables: confirmed members of the catalog (Table~\ref{tab:YMDB_confirmed}), candidates pending spectral classification (Table~\ref{tab:YMDB_candidates}), and unqualified candidates (Table~\ref{tab:YMDB_unqualified}). Confirmed members are systems that meet the essential conditions of displaying Algol-type LCs (detached) and having at least one component with a spectral classification within the O9-B1 V range. Candidates are also detached systems that show potential signs of meeting the spectral type criteria but lack definitive confirmation due to incomplete or inconclusive spectral classification. Unqualified systems are those whose LCs do not appear detached (EB or EW), have primary spectral classifications clearly out of range, or both. The secondary component's classification might either be out of range or impossible to detect due to limitations in current spectroscopic data and analysis. While unqualified detached systems with indeterminate secondary classifications are currently unsuitable for the catalog, future higher-resolution or more sensitive spectra could reveal additional stellar features, prompting their reconsideration as candidates.

For each group, we detail findings from our photometric analysis, including delta magnitude ($\Delta mag$), period ($P$), time of minima ($T_0$), observed apsidal motion (Apsidal), additional system variability (MultiP), heartbeat phenomena (HB), and eccentricity ($e$). Additionally, we provide our spectral classifications for certain stars, either to fill gaps where classifications were absent or to verify existing classifications.

Regarding new spectral types, we confirmed HD~298448, CD-28~5257, HD~309036, V*~IK~Vel as entries to the YMDB (Table~\ref{tab:YMDB_confirmed}),
and emphasize that certain eclipsing binary systems were excluded from the YMDB catalog because their spectral types fall outside the range studied here. 
An intriguing example is CM CMa (=Gaia DR3 2928505622380096256), which lacks a documented spectral type in the eight publications indexed in the Simbad database. Our April 2023 night spectrum reveals absorptions consistent with an F5 V type, characterized by similar H and K Calcium lines and the presence of the G-band.
Another case is GN Nor (=Gaia DR3 5884730512723833984), which similarly lacks spectroscopical references in Simbad. Our CASLEO spectrum is classified as A0 V, primarily due to the dominance of Balmer lines.
Similar cases are V*~GN~Nor and *~f~Vel, whose spectral classifications resulted out of the range for this catalog.
These spectra are shown in Fig.~\ref{fig:espectros}.

In certain LCs, we observed additional variability beyond the typical eclipsing patterns. These variations, unless identified as heartbeat phenomena, are denoted with 1 in the tables. Therefore, we detected such oscillations in 37 systems, 11 of which are unprecedented.
Most newly discovered pulsating systems have dedicated paragraphs, except 
HD~309036, 
HD~204827, 
all of them well known DEBs. 

Finally, the visual inspection of TESS LCs allowed us to detect apsidal motion. This phenomena is indicated in the tables (with 1) when the times of the secondary eclipses seem to vary among different sectors.

\subsection{New detached eclipsing binaries discovered} \label{subsec:NewDEB_discovered}

\begin{table}
    \centering
    \caption{Newly identified eclipsing binaries.}
    \label{tab:starlist_v2}
    \begin{tabular}{lccr}

\hline
\hline
Simbad          & RA            & Dec           & V             \\
\hline
\multicolumn{4}{c}{New variables discovered (DEB)}              \\
\hline
BD+66  1674     & 00 02 10.2414 & +67 25 45.186 & 9.6           \\
BD+66  1675     & 00 02 10.2887 & +67 24 32.228 & 9.08          \\
HD 278236       & 05 26 55.2050 & +40 32 53.088 & 10.86         \\
RAFGL 5223      & 07 08 38.7906 & -04 19 04.847 & 12.06         \\
TYC 8174-540-1  & 09 20 18.7707 & -49 50 25.869 & 11.63         \\
HD 102475       & 11 47 18.1823 & -62 26 10.246 & 9.09          \\
HD 114026       & 13 08 44.0123 & -60 20 18.400 & 9.36          \\
CPD-63  3284    & 14 31 58.7177 & -63 36 17.783 & 11.2          \\
UCAC2   5911156 & 15 15 25.1420 & -59 09 48.929 & 12.17         \\
HD 142152       & 15 55 33.8521 & -54 46 35.833 & 9.82          \\
CD-54  6456     & 15 55 39.6014 & -54 38 36.636 & 10.41         \\
HD 144918       & 16 10 29.1581 & -49 02 47.091 & 9.96          \\
BD+55  2722     & 22 18 58.6254 & +56 07 23.482 & 10.15         \\
\hline
\multicolumn{4}{c}{New eclipsing binaries discovered (DEB)}     \\
\hline
HD 277878       & 05 18 22.7563 & +41 56 06.000 & 10.27         \\
* psi02 Ori     & 05 26 50.2293 & +03 05 44.422 & 4.61          \\
LS  VI +00   25 & 06 48 50.4782 & +00 22 37.694 & 10.86         \\
CD-53  6352     & 16 00 26.8255 & -53 54 39.860 & 10.4          \\
HD 338961       & 19 50 24.4190 & +27 27 55.913 & 10.86         \\
\hline
\multicolumn{4}{c}{New eclipsing binaries discovered (not DEB)} \\
\hline
LS   V +38   12 & 05 20 00.6489 & +38 54 43.505 & 10.4          \\
HD 305850       & 11 01 52.2783 & -60 00 46.795 & 8.8           \\
\hline
\hline

    \end{tabular}
    \begin{justify}
        \textbf{Note:} List of 20 eclipsing binaries identified in this study. This includes 13 binaries previously unrecognized as variables, five known either as photometric variables or spectroscopic binaries but not as DEBs, and two identified as nonthermal contact binaries.
    \end{justify}      
    
\end{table}
We offer a detailed insight into select DEB featured within the YMDB.

\begin{figure}
    \centering
    \includegraphics[width=1.0\columnwidth]{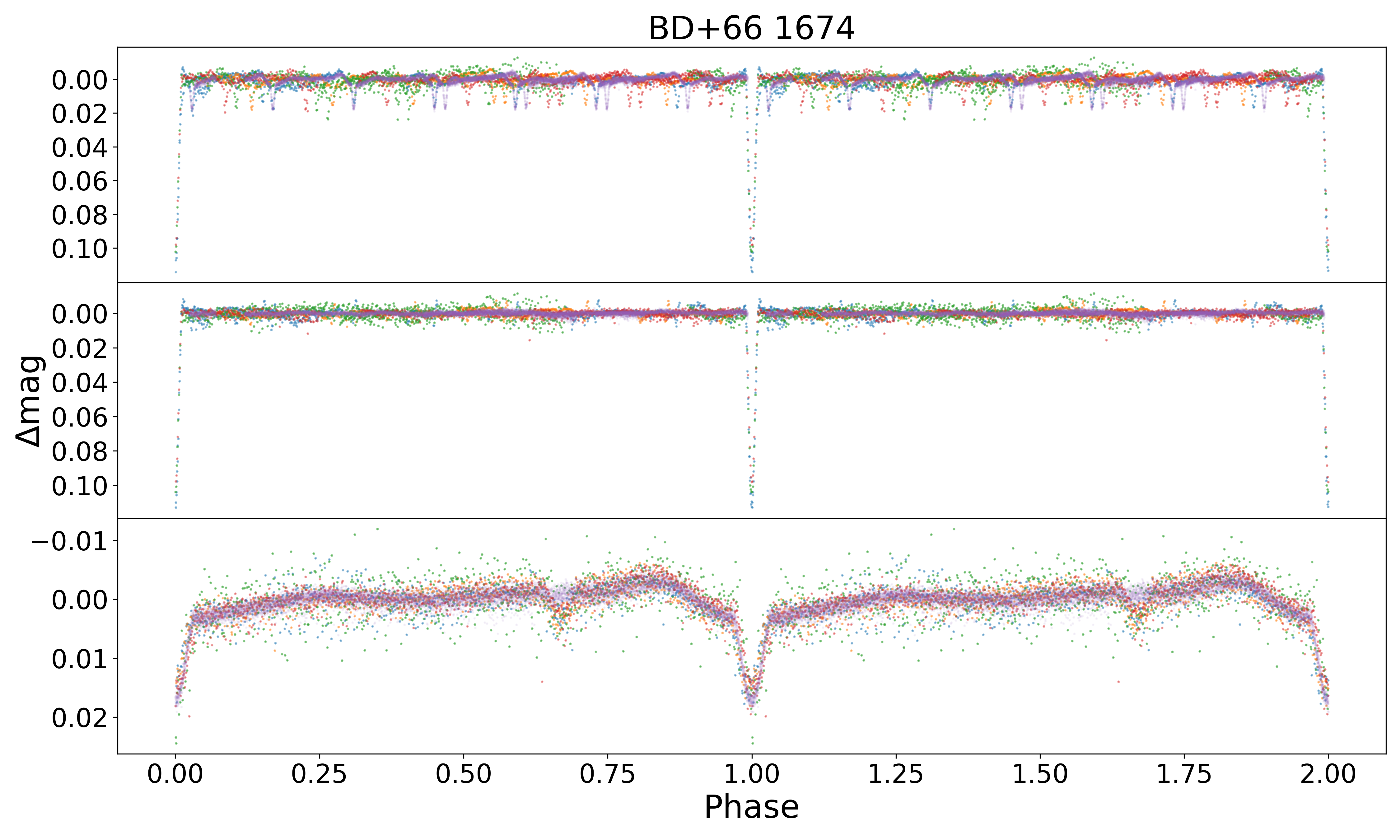} 
    \caption{Light curves of BD+66 1674. The top panel displays the composite LC, while the middle panel shows the LC corresponding to the B0 V+ B0 V system. In the bottom panel, the LC of the newly discovered eccentric short-period DEB system is depicted, with its sources unidentified as of yet. Each color represents a different sector of TESS.}
    \label{lc:BD+66 1674J}
\end{figure}

\textbf{BD+66 1674}: 
Initially noted as a probable RV variable \citep{1974PDAO...14..283C}, no additional references were found in the available literature. Although \citet{2016ApJS..224....4M} classified it as O9.7 IV, an analysis, incorporating high-resolution spectra from the Galactic O-Star Spectroscopic Survey \citep[GOSSS;][]{2016ApJS..224....4M} and the Library of Libraries of Massive-Star High-Resolution Spectra \citep[LiLiMaRlin;][]{2019hsax.conf..420M}, revealed it to be a B0 V + B0 V system (Maíz Apellániz, priv. comm.). Consequently, it has been included in the YMDB.  Its LC displays additional variations beyond the 18-day eclipsing pattern, resembling those of highly eccentric detached double-eclipsing systems. Moreover, a notable bump between eclipses suggests the presence of a heartbeat-like feature (see Fig.~\ref{lc:BD+66 1674J}). Considering the pixel size of TESS data, we conducted a search for other targets within a 40 arcsec radius and identified seven Gaia sources. These sources, at least four G-magnitudes fainter than BD+66 1674, complicate the determination of whether the short-period binary is one of these sources or an unresolved component of BD+66 1674.  

\begin{figure}
	\includegraphics[width=1.0\columnwidth]{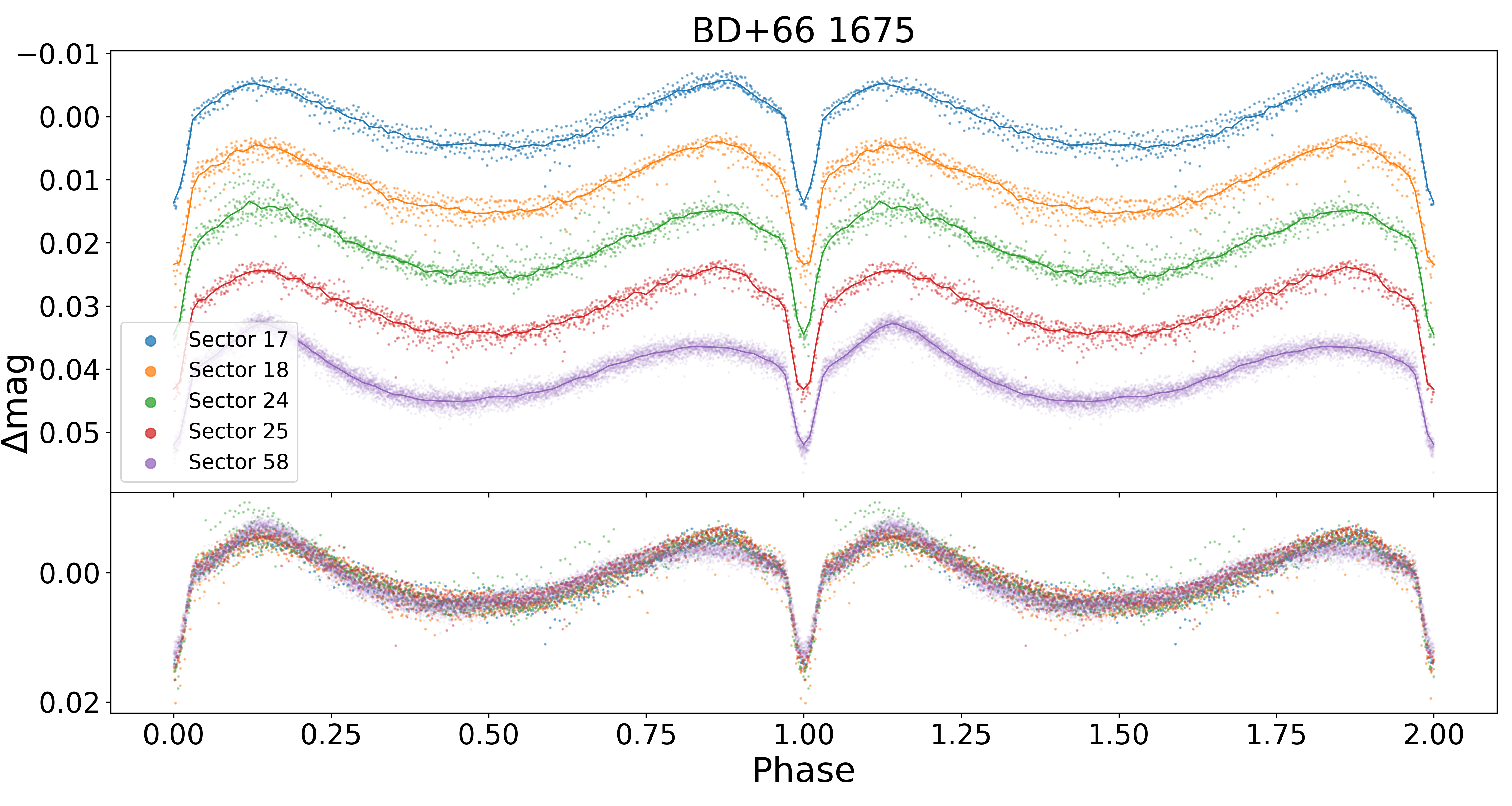} 
	\caption{Light curves of BD+66 1675. The top panel displays each reduced TESS sector separated by an artificial 0.01 $\Delta$ Mag for visibility. The composite LC is shown in the bottom panel. Sectors from 17 to 58 span 1150 days of observation. A heartbeat-like feature is present before the eclipse in the earliest sector (17), while the same feature is visible after the eclipse by the latest sector (58).}
	\label{lc:BD+66 1675_E}
\end{figure}

\textbf{BD +66 1675}: 
Classified as O7.5 Vz \citep{2016ApJS..224....4M}, high-resolution spectra obtained within the context of LiLiMaRlin \citep{2019hsax.conf..420M} revealed BD +66 1675 to be a triple system comprised of O7.5Vz+O8V~+B components. A preliminary analysis of 20 spectra indicated that the RV of the spectral lines belonging to the O7.5~Vz star do not vary within errors. However, the O8V~+~B pair exhibits RV motion in accordance with the photometric period. Moreover, preliminary orbital elements suggest an eccentric orbit, explaining why only one eclipse is observed. This eclipse occurs when the massive star passes in front of the system, making it a secondary eclipse. Also, we detect a heartbeat-like behavior immediately after the eclipses, but only in the Sector 58 (Fig. ~\ref{lc:BD+66 1675_E}), and short-period variability identified as pulsations.

\begin{figure}
	\includegraphics[width=1.0\columnwidth]{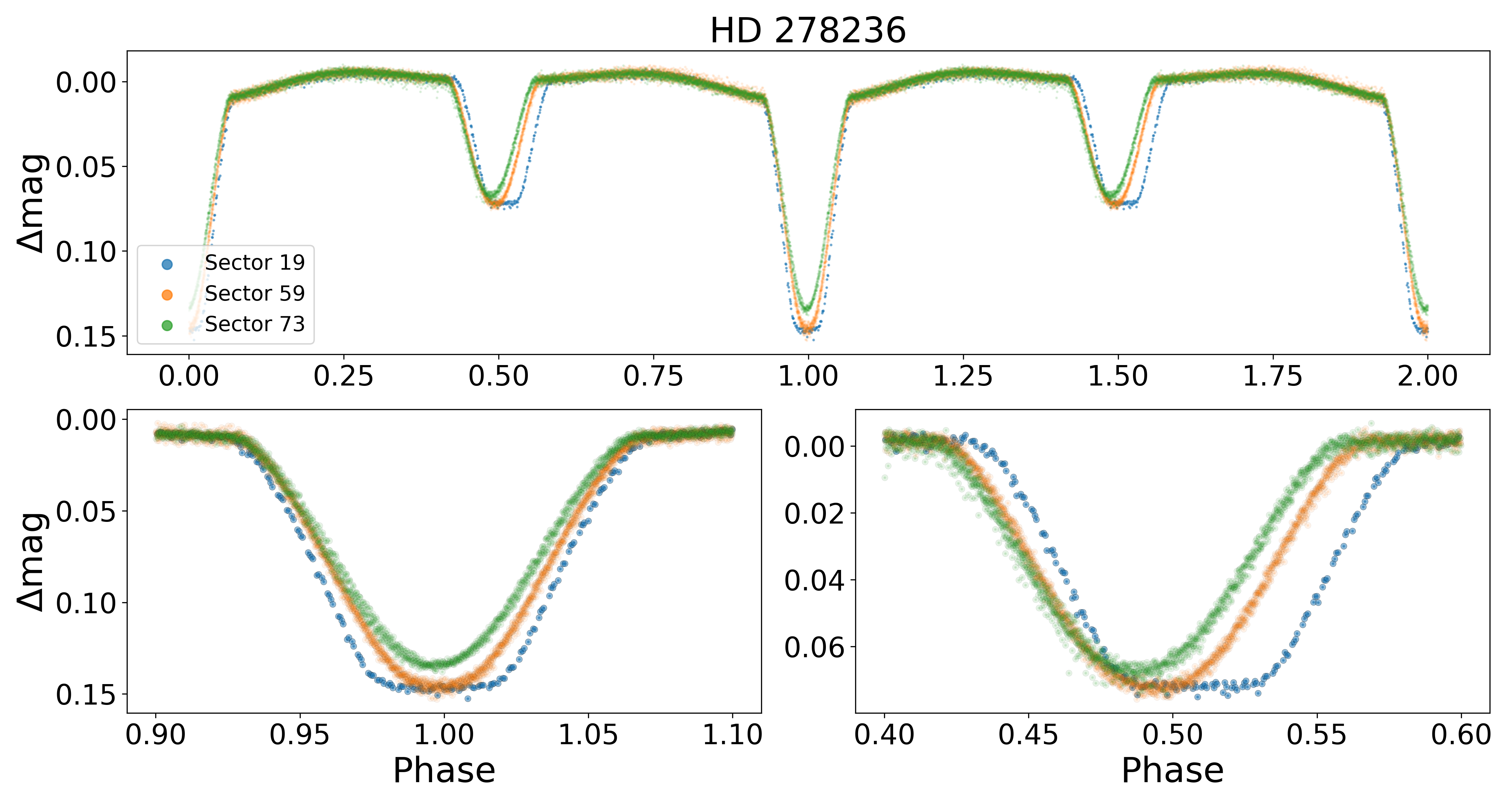} 
	\caption{Light curves of HD 278236. The top panel displays the composite LC of the system. Bottom left and right panels provide zoomed views of the primary and secondary eclipses, respectively. Observations from Sectors 19 to 73 span 1500 days. In the earliest sector (17) the eclipse exhibits a flat profile, indicative of a total eclipse. Over time the eclipse profile gradually smoothens, completely losing its flatness by the latest observed sector (73), indicative of a shift to a partial eclipse. Apsidal motion is also evident in this system, as demonstrated by the phase shift of the secondary eclipse.}
	\label{lc:HD 278236}
\end{figure}

\textbf{HD 278236}: 
Classified as O9 V by \citet{1973AnA....25..337G}, no publications reporting variability or binarity were found. It is noteworthy that the relative position of the secondary eclipse appears to vary in different sectors, indicating apsidal motion (Fig.~\ref{lc:HD 278236}). Moreover, apsidal motion is making the secondary eclipse narrow overtime (from sector 17-59 $\sim$1100 days and from 59-73 equivalent to $\sim$1400 days, for a total time span of $\sim$2500 days).

\textbf{RAFGL 5223}: 
Recognized as a Herbig Ae/Be star candidate, its entire bibliography is centered around this characteristic. However, no indications of variability, either spectroscopic or photometric, were identified. To scrutinize the indicated spectral classification in the Simbad database, we downloaded a X-Shooter spectrum (program ID 084.C-0952(A)) from the ESO portal. By evaluating the ratios between He \textsc{ii} and He \textsc{i} pairs, we determined it to be of O9.2 type, with He \textsc{ii} $\lambda$4686/He \textsc{i} $\lambda$4713 and Si \textsc{iv} $\lambda$4089/He \textsc{i} $\lambda$4026 ratios indicative of a V luminosity class. Consequently, it has been included in YMDB as an eccentric short-period binary. We note a potential heartbeat feature, manifested as a bump in the orbital phases where the periastron passage is expected.
We also remark a probable apsidal motion, as separation between eclipses seems to vary among Sectors 7 and 33 ($\sim$700~d).

\textbf{TYC 8174-540-1}: 
Initially identified as an OB star by \citet{1977AJ.....82..474M} and subsequently classified as O9.5 Vn by \citet{1982MNRAS.201..885B}, TYC 8174-540-1's spectral type was verified through our own CASLEO spectrum (see Fig.~\ref{fig:espectros}). Despite an absence of reports regarding its binary nature, the TESS LC unmistakably depicts both eclipses, which, in fact, are transit and occultation events. This characteristic increases its significance as a benchmark for determining stellar parameters. Notably, the system exhibits eccentricity and appears to manifest a heartbeat phenomenon before entering the primary eclipse.

\textbf{HD 102475}: 
We found no evidence in the existing literature suggesting that it is a binary or variable star. It is only mentioned in general works on spectral types \citep{1961MNRAS.122..239F,1975mcts.book.....H}, where it is classified as B0.5 II and B1 III, respectively. Therefore, we include this target as a candidate until its spectral type is confirmed with modern spectroscopic analysis. It also exhibits short-period variability, interpreted as pulsations.

\textbf{HD 114026}: 
While we found no evidence suggesting variability, \citet{1983ApJS...52....1G} identified it as a probable SB2 system. Despite its spectral classification exhibiting significant dispersion, ranging from OB to B2, with \citet{1983ApJS...52....1G} designating it as a B0.5 V:n type, we prefer to be cautious and include HD~114026 as a candidate for the YMDB.

\textbf{CPD $-$63 3284}: 
This star lacks a reliable spectral classification, and unfortunately, we were unable to observe it during our run at CASLEO. Initially identified as an OB star by \citet{1964MeLuS.141....1L}, no additional classification information was found. Nevertheless, its LC unmistakably displays double eclipses, which indeed correspond to occultation and transit events. It also exhibits short-period variability, interpreted as pulsations. We include it as a candidate in the YMDB.

\begin{figure*}
	\includegraphics[width=2\columnwidth]{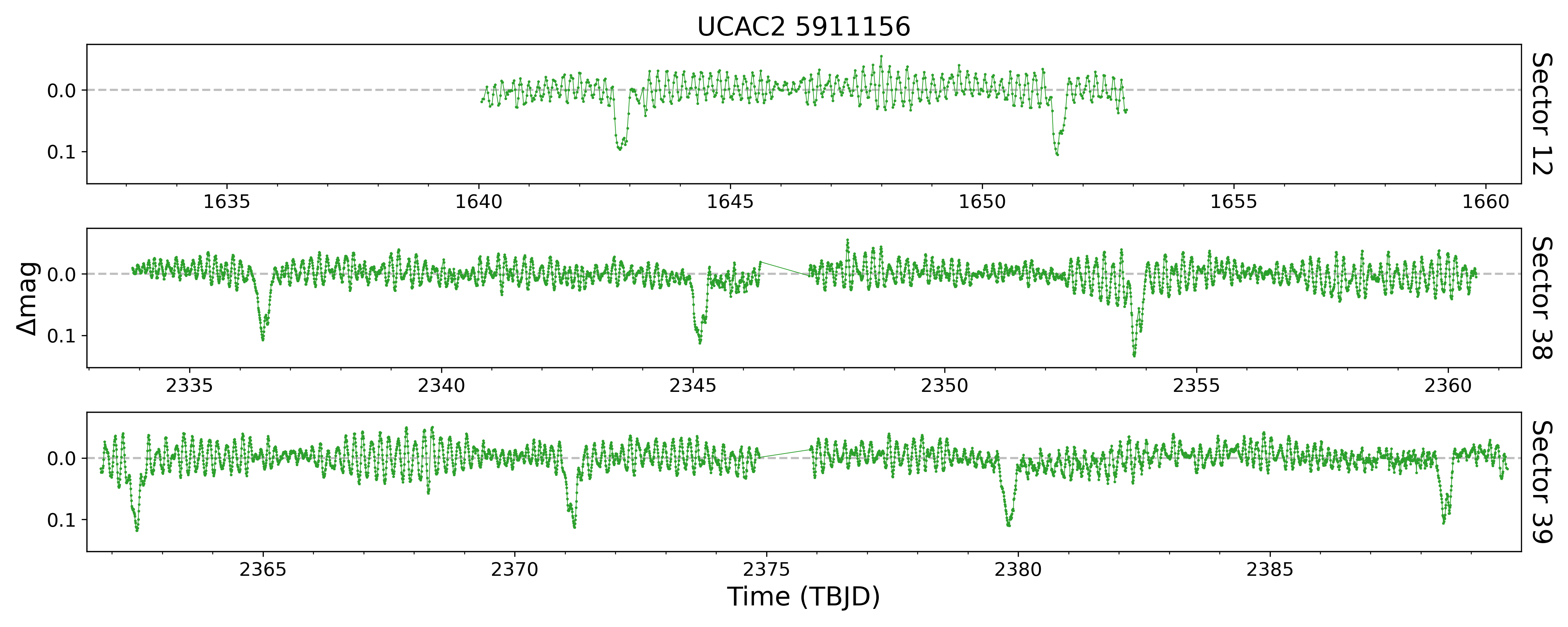} 
	\caption{Light curves of UCAC2 5911156. Panels top to bottom show the unfolded LCs of Sectors 12, 38, and 39 respectively. All panels maintain equal Y and X axis scales. This system exhibits pronounced Tidally Excited Oscillations (TEOs) in each sector, reflecting oscillations in the star or system caused by tidal interactions with a nearby stellar companion.}
	\label{lc:UCAC2 5911156 unfolded}
\end{figure*}

\textbf{UCAC2 5911156}: 
Initially recognized as an early-type star by \citet{1977AJ.....82..474M} and subsequently classified as B0.5 V \citep{1982MNRAS.201..885B}, we acquired a new spectrum leading to a reclassification of its luminosity class to III. This adjustment is based on the almost similar intensities of the absorption lines He \textsc{i} $\lambda$4387 and Si \textsc{iii} $\lambda$4552. The giant nature of UCAC2~5911156 naturally explains the pulsations noted in the LC shown in Fig.\ref{lc:UCAC2 5911156 unfolded}. The LC exhibits eclipses with a superimposed set of oscillations forming a beating pattern, akin to those initially discovered in HD187091 \citep[][and references therein]{2011ApJS..197....4W,2012ApJ...753...86T}. Notably, these damped oscillations do not align with the orbital period, as the maximum amplitudes do not occur at the same phase. To the best of our knowledge, this is the first identification of the system as a DEB and pulsating. However, a detailed analysis of this intriguing system is beyond the scope of this work. 

\textbf{HD 142152}: 
Initially classified as a B0 III massive star by \citet{1971AJ.....76..260C} and subsequently confirmed by \citet{1975mcts.book.....H}, HD 142152 lacked published spectra. To address this absence, we observed this target using CASLEO and confirmed its spectral classification (see Fig.~\ref{fig:espectros}). The identification as a probable binary arises from two discrepant radial velocities reported by \citet{1972MNRAS.158...85C}. The TESS LC distinctly illustrates a highly eccentric detached double-eclipsing behavior. Additionally, pulsation-like variations, occurring with 1/8 of the orbital periodicity, and a discernible heartbeat feature are detected. This renders the target exceptionally interesting, despite its exclusion from YMDB due to its luminosity class.

\textbf{CD $-$54 6456}: 
Prior to this study, there were no records indicating variability or binarity for this system. Classified as O9.5 V \citep{2014ApJS..211...10S}, our analysis involved six high-resolution spectra, which revealed no RV variations in the features of the O-type star. Consequently, this suggests that the eclipsing binary likely involves another stellar pair within the TESS pixel or is indistinguishable from CD $-$54 6456. For this reason, we have included it as a candidate in the YMDB.

\textbf{HD 144918}: 
This target exhibits one of the most widely scattered spectral type assignments in the literature, ranging from O5/7 \citep{1978mcts.book.....H} to B0 \citep{1921AnHar..96....1C}, with none originating from a modern study. Consequently, we decided to acquire a new spectrum. The CASLEO spectrum is classified as O8 V, with this classification based on the observation that He \textsc{ii} $\lambda$4542 is slightly fainter than He \textsc{i} $\lambda$4471. Concerning its binary nature, it is reported as SB2 by \citet{1963MmRAS..68..173F}, yet no period is provided, and no dedicated work was found in the literature. 

\textbf{BD +55 2722}: 
This system is, indeed, a Trapezium-like configuration with components named A (O8 Vz), B (O9.5 V), and C (O7 V(n)z+B), according to the GOSSS \citep{2016ApJS..224....4M}. Given that the three sources are indistinguishable in TESS data, we attribute the DEB discovery to the C component, thereby designating BD +55 2722~C as a candidate in the YMDB. This classification stands until the spectral type of the secondary component is more precisely determined.  

\textbf{HD 277878}: 
Despite extensive literature search, no evidence of photometric variability was found for this target. Originally identified as an OB or B0-type star, it was recently reclassified as O7 V((f))z based on LAMOST spectra \citep{2021ApJS..253...54L}, and indicated as SB1. Consequently, we have excluded it from consideration in the YMDB.

\textbf{* psi02 Ori}: 
It is a widely recognized spectroscopic binary \citep{1908ApJ....28..266P,1985PASP...97..428L}, and also known to exhibit ellipsoidal variations \citep{2013AN....334..860A}. In the TESS data, its double-eclipsing nature is clearly evident, alongside the ellipsoidal variations. However, due to its spectral type being identified as B1 III + B2 V \citep{1985PASP...97..428L}, we have excluded it from consideration in the YMDB.

\textbf{LS  VI +00   25}: 
It was initially identified as an SB1 system by \citet{1999A&A...343..806M}, although no photometric variations have been reported to our knowledge based on available literature. With a spectral type of O9.5 V \citep{2021ApJS..253...54L}, we have included it in the YMDB.

\textbf{CD-53 6352}: 
This star is recognized as double or multiple according to the Washington Double Star Catalog (WDS J16001-5355AB), but no photometric or RV variability has been identified. Analysis of TESS data has not conclusively determined whether star A or B is the DEB. While component A is classified as O7 III \citep{1993ApJS...89..293V}, the spectral classification for component B is unavailable. As such, we have included it as a candidate in the YMDB. Additionally, its LC exhibits short-period variations compatible with pulsations.

\textbf{HD 338961}: 
No evidence of variability or binary nature was found for this star in the literature consulted. Despite some dispersion in its spectral classification, we consider the determination B0.5 IIInn from \citep{1980ApJ...235..146T} to be reliable. Therefore, we have excluded it from inclusion in the YMDB.

\subsection{Contact binary systems discovered}
We provide details of two targets initially considered for inclusion in the YMDB. However, subsequent analysis of the TESS data revealed that they are contact binaries, leading to their rejection.

\textbf{LS V +38 12}: 
This star was identified as a binary system, based on spectroscopic observations, by \citet{2016ApJS..224....4M}, who classified the pair as O7 V((f))+ B0 III-V.

\textbf{HD 305850}: 
It is listed as a pulsating star in Simbad, yet we found no references to its periodicity or LC.  Moreover, its spectral type is actually uncertain, requiring an accurate determination.

\subsection{DEBs showing heartbeat phenomena}  \label{heartbeat}

In the YMDB, in addition to TYC 8174-540-1, RAFGL 5223, HD 142152, BD+66 1675 and the subsystem found within BD+66 1674  which are already detailed in \ref{subsec:NewDEB_discovered}, several other systems exhibit variations in their LCs consistent with the heartbeat phenomenon. 

\textbf{V* NY Cep}: 
It is a well-known DEB \citep[see e.g.,][]{2011ApJ...726...68A}, characterized by a single eclipse resulting from its high eccentricity and periastron longitude. Notably, a bump is observed in its LC just before entering the eclipse, which we interpret as a probable heartbeat, although other proximity effects could not be discarded. 

\textbf{HD 99898}: 
This star is a previously identified variable, classified as Algol type \citep{1998AcA....48...35P}, and subsequently analyzed as a detached eclipsing binary (DEB) displaying apsidal motion \citep{2006AstL...32..772K}. However, no heartbeat phenomenon was reported prior to this study.

\textbf{CD-28 5257}: 
It is a recognized eclipsing binary \citep[see e.g.,][]{2019MNRAS.490.5147P}, although the reported periodicity is incorrect. Our analysis reveals it to be an eccentric double-eclipsing system, with eclipses corresponding to both transits and occultations. Furthermore, a marginal bump is observed in the orbital phases where the expected periastron occurs, which we identify as a heartbeat effect, made possible by the precision of the TESS data. This star is marked for exhibiting apsidal motion in Table~\ref{tab:YMDB_confirmed}, as the orbital phases of the secondary transits appear to vary across available Sectors 7, 8, 34, and 61.

\textbf{V* V725 Car}: 
It is a DEB in a highly eccentric orbit, with an RV curve determined for its primary component \citep{2018MNRAS.477.2068K}. The new LC presented here reveals a distinct heartbeat effect between the closer eclipses. An integrated analysis of both datasets, spectroscopic and photometric, will enable the determination of stellar parameters for both components.

\textbf{V* Y Cyg}: 
It is a well-established SB2 eclipsing binary \citep{1887AJ......7..116S,2019AnA...626A..20M}. A marginal flux increase is observed at orbital phases corresponding to the periastron passage, which can be interpreted as a heartbeat effect, although it could also be attributed to ellipsoidal variations. This system exhibits apsidal motion, as evidenced by the variation in secondary eclipse times (and possibly their depths) between sectors 15 and 41 ($\sim$700 d).

\textbf{CD -27 4726}: 
It is a very well-known DEB. TESS data show an asymmetric light increasing just before and after the (unique visible) primary eclipse. Its shape reminiscences the one detected on the very massive binary system WR~21a \citep{2022MNRAS.516.1149B}. Unluckily, the spectral type of this DEB is not completely reliable thus, we add CD~$-$27~4726 to the candidates list.

\textbf{HD 306096}: 
It is also a very studied DEB, but we note a clear heartbeat feature on the orbital phases where the periastron passage is expected.

\textbf{V* GN Nor}: 
This is another well-known system; however, its heartbeat has not been reported to the best of our knowledge.

\textbf{*f Vel}: 
It is a studied DEB system; however, as far as we are aware, its heartbeat phenomenon has not been documented.

\subsection{DEBs with newly determined periods}  \label{subsec:NewDEB_periods}
We compared the periods obtained from our analysis of the TESS data with those previously published. Some of the periods we derived were found to be twice as long as the previously reported ones. This discrepancy is likely due to the improved depth and resolution of the eclipses identified in the TESS data, enabling a clearer distinction between primary and secondary eclipses. While we have already discussed these discrepancies for system CD-28 5257 in Section \ref{heartbeat}, we now address those for which we have not yet provided details.

\textbf{V* KU Car}: 
KU Car, reported as B8 in SIMBAD from \citet{2013AN....334..860A} and initially observed with a period of 5.92 days by \citet{1951PRCO....2...85O}, is reexamined in the author's Master's dissertation \citet{2021mobs.confE..36M}. The dissertation utilized ASAS-3 data, which revealed a secondary eclipse previously unnoticed, effectively revising the reported period to 2.96 days. Detailed spectral analysis was performed using data from the Galactic O Star Spectroscopic Survey \citet[GOSSS;][]{2011ApJS..193...24S,2014ApJS..211...10S,2016ApJS..224....4M}, which led to a revised spectral classification of B0.5 V(n). While O'Connell previously suggested an eccentric orbit based on the LC's shape outside the eclipses, this hypothesis was biased by an incomplete understanding of the LC's true nature. Our observations do not indicate eccentricity from the LC; only a comprehensive RV study could resolve this ambiguity. Ongoing spectroscopic analysis aims to refine the temperature and absolute parameters of KU Car.

\textbf{HD 99630}: 
This star is known as DEB, but its periodicity is badly reported in the literature \citep{1998AcA....48...35P,2012A&A...548A..79A}. Our analysis of the TESS LC reveals that its period is nearly double the previously reported value. Additionally, it shows double eclipses and high eccentricity. Our high-resolution spectra, obtained during the OWN Survey campaigns, indicate its spectral type is earlier than the B4-B5 determined by \citet{1976AnAS...23..283L}. The ratio between He \textsc{i} $\lambda4471$ and Mg \textsc{ii} $\lambda4481$ lines is greater than three, and C \textsc{ii} $\lambda4267$ is identified, thus a B3-type is more suitable. The weakness of the metal lines confirms its dwarf class, B3 V.

\begin{figure}
	\includegraphics[width=1.0\columnwidth]{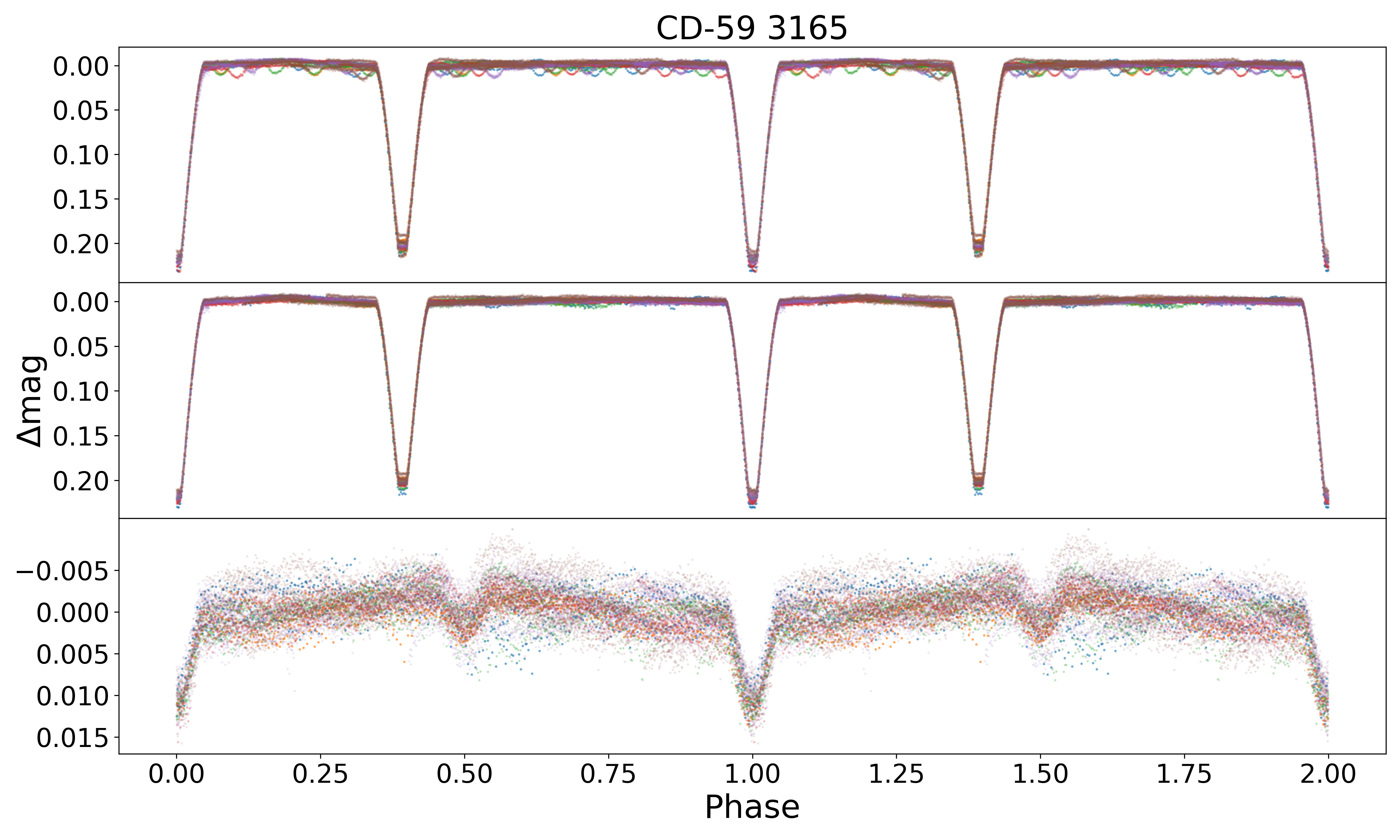} 
	\caption{Light curves of CD-59 3165. The top panel displays the composite LC. The middle panel illustrates the LC for the system with a period of ~7.59 days, while the bottom pannel shows the system with a period of ~3.18 days. Each color represents a different sector of TESS}
	\label{lc:cd-59_3165J}
\end{figure}

\begin{figure}
	\includegraphics[width=1\columnwidth]{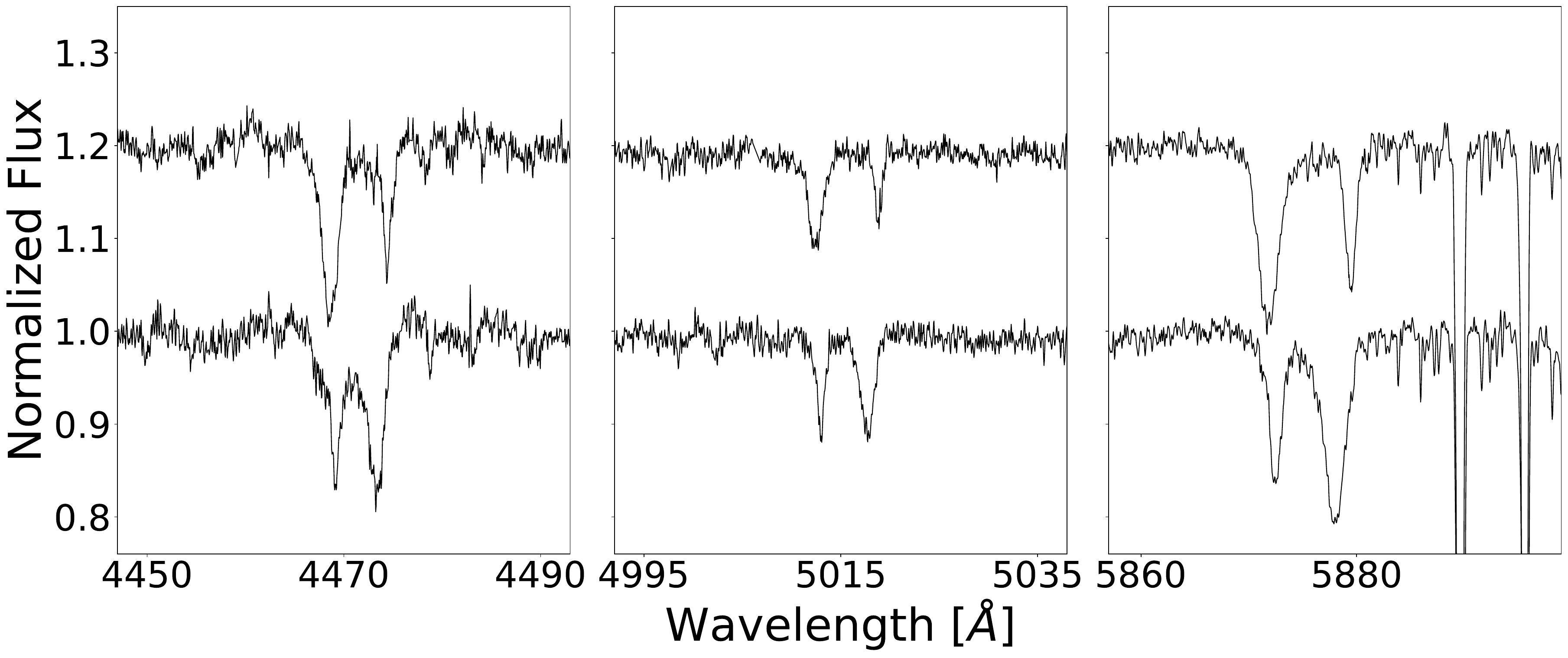} 
	\caption{Three wavelength regions, around selected He \textsc{i} absorption lines of two spectra of CD -59 3165 obtained during quadratures.}
	\label{sp:cd-59_3165}
\end{figure}

\textbf{CD-59 3165}: 
It is recognized as a highly eccentric DEB \citep{2018ApJS..235...41K}. In the TESS data, in addition to exhibiting double-eclipsing behavior, it also displays other eclipses with a periodicity of 3.18205~d (see Fig.~\ref{lc:cd-59_3165J}). Given that this star is identified as a double in the WDS catalog (WDS J10348-6013AB), with both stars separated by only 2.3 arcseconds \citep{2001AJ....122.3466M}, we can not definitively confirm the origin of these additional eclipses. However, our analysis of two spectra obtained during opposite quadratures of the main DEB system as part of the OWN Survey program reveals distinct spectral features for each component. One component exhibits narrow lines, while the other shows broader lines. The ratio of Si \textsc{iii} $\lambda$4552 to Si \textsc{iv} $\lambda$4089 is nearly unity in the narrow component, suggesting a spectral type of B0.5. Conversely, this ratio is smaller in the broader component, and considering the absence of He \textsc{ii} $\lambda$4542, yield a spectral type of B0-0.2. In terms of luminosity class, the narrow component appears to belong to classes III-I, as He \textsc{ii} $\lambda$4686 is markedly fainter than He \textsc{i} $\lambda$4713. Conversely, both lines are comparable in the broad component, indicative of a class V classification. In Fig.~\ref{sp:cd-59_3165} we show some spectral regions to illustrate these classifications.
In the future, we plan to conduct a more targeted spectral analysis to identify the spectral signatures of both components of the other DEB system. This analysis will also aim to elucidate the source of the additional eclipses observed.

\subsection{Other intriguing systems}
We identified a diverse set of interesting systems, each presenting unique photometric characteristics that merit special mention. Firstly, several systems in our sample displayed Tidally Excited Oscillations (TEOs), which are oscillations within a star or system driven by tidal forces due to a close stellar companion. Notable examples include UCAC2 5911156 (detailed in \ref{subsec:NewDEB_discovered}, Fig.~\ref{lc:UCAC2 5911156 unfolded}) along with V* V4386 Sgr, 2MASS J16542949-4139149, * 23 Ori, and V* V1216 Sco.

\begin{figure}
	\includegraphics[width=1.0\columnwidth]{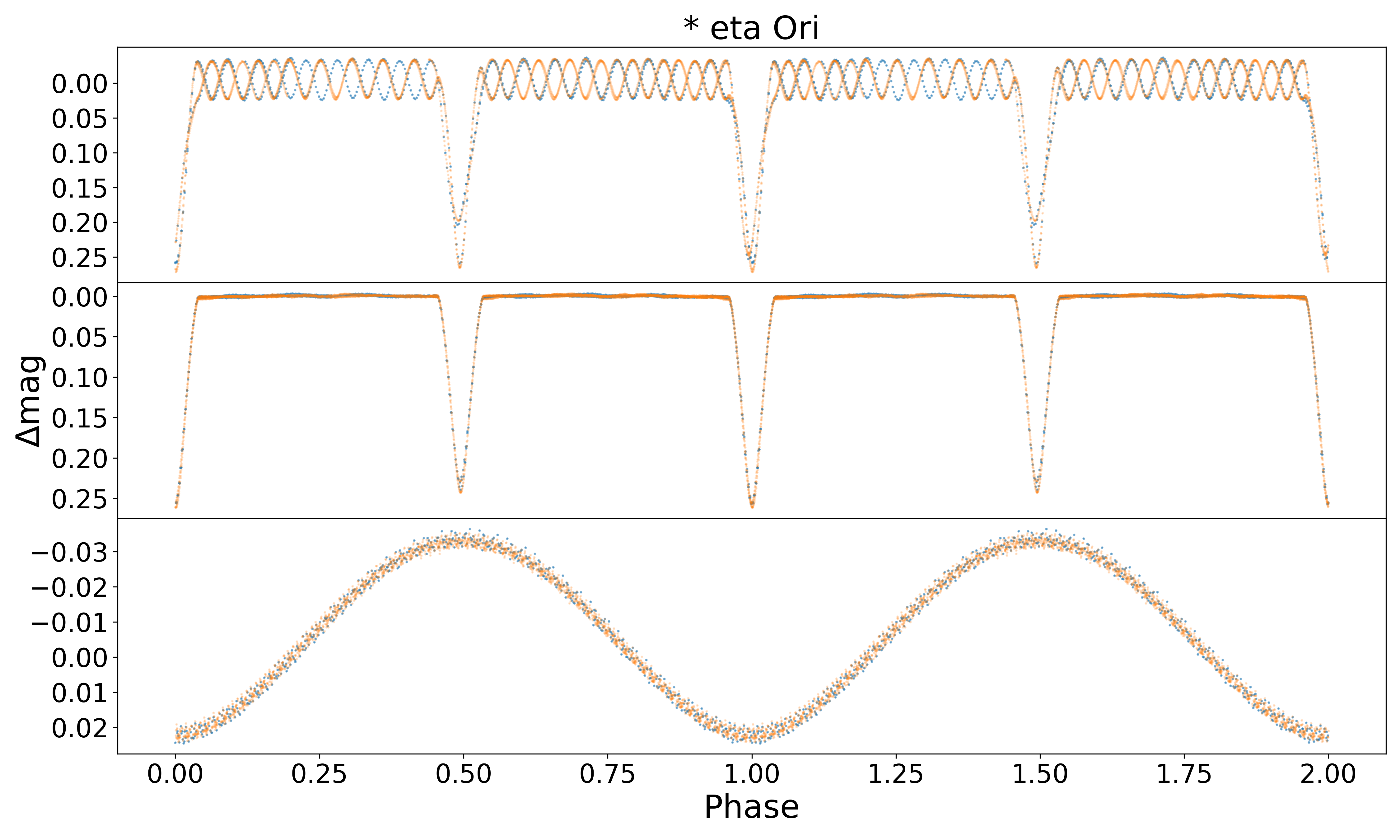} 
	\caption{Light curves of * eta Ori. The top panel displays the composite LC. The middle panel illustrates the LC for the eclipsing system with a period of ~7.989 days disentangled from the shorter period pulsations, while the bottom pannel shows such pulsations folded with period of ~0.432 days. Each color represents a different sector of TESS}
	\label{lc:eta_OriJ}
\end{figure}

Additionally, we have successfully disentangled LCs of systems where the combined photometric data initially obscured individual components. Systems such as CD-59 3165 (detailed in \ref{subsec:NewDEB_periods}, Fig.~\ref{lc:cd-59_3165J}), BD+66 1674 (\ref{subsec:NewDEB_discovered}, Fig.~\ref{lc:BD+66 1674J}) and eta Ori (detailed below, Fig.~\ref{lc:eta_OriJ}). 

\begin{figure*}
	\includegraphics[width=2\columnwidth]{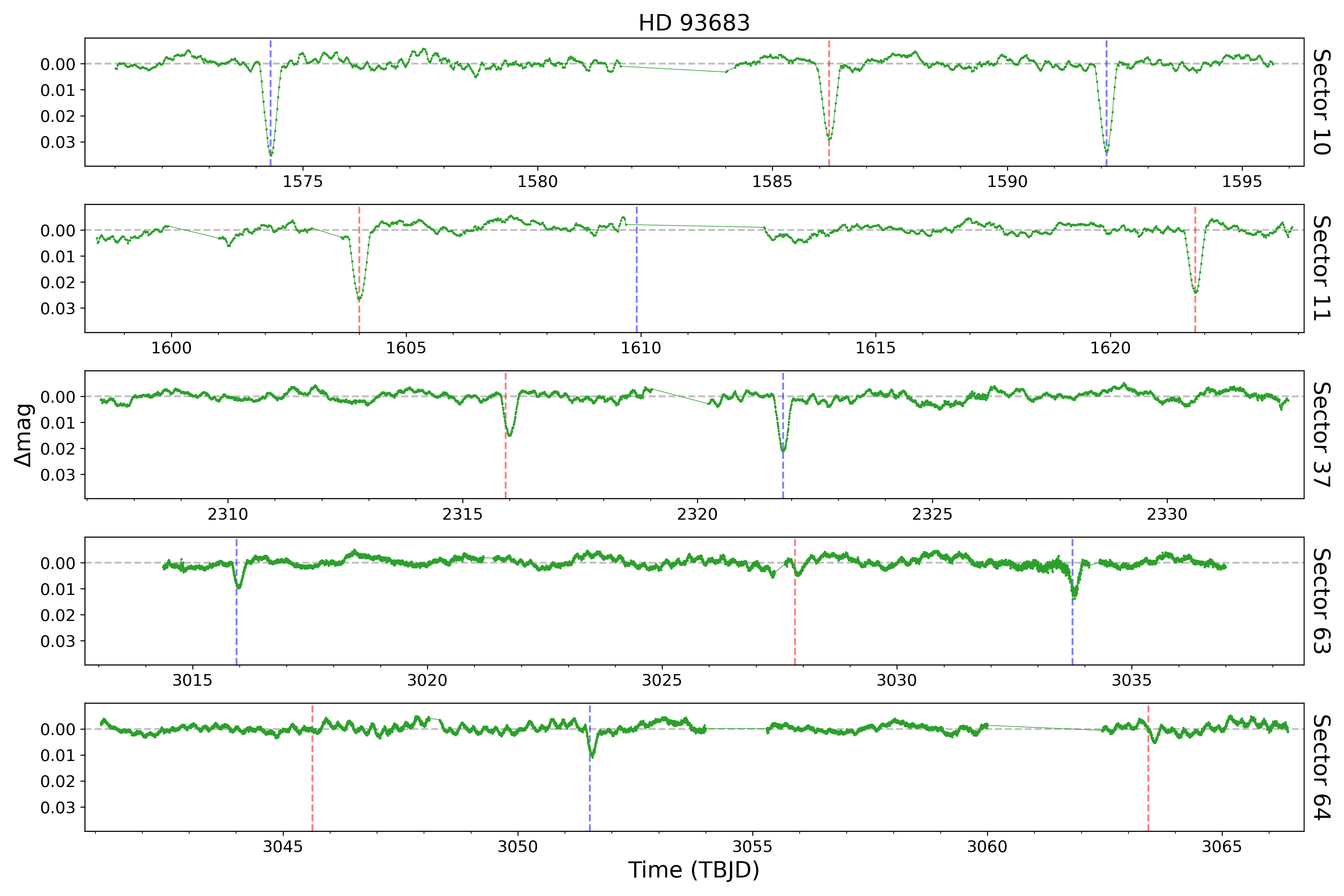} 
	\caption{Light curves of HD 93683. Each panel represents a different TESS sector, arranged chronologically from top to bottom. Predicted timings for the primary and secondary eclipses are marked with blue and red dashed vertical lines, respectively. All panels maintain equal Y and X axis scales. Over ~1500 days, both primary and secondary eclipses consistently decrease in depth from approximately 0.3$\Delta$mag to around 0.1$\Delta$mag, with the secondary eclipses being shallower overall. By the latest sector (64), the depth of the eclipses approaches the level of the system's intrinsic variability, rendering the first secondary eclipse undetectable. Apsidal motion is evident as early as Sector 37 when compared to Sector 10, although data from later sectors are not reliable for further apsidal motion analysis.}
	\label{lc:HD_93683 unfolded}
\end{figure*}

We also observed systems whose eclipse characteristics vary significantly over time, adding another layer of complexity to their study. For instance, HD 278236 initially showed flat eclipses typical of total eclipses but over a span of 1500 days, the eclipse profile gradually transitioned to partial, with a smoothing and narrowing that indicates dynamic changes in the system. Additionally, the shifting of the secondary eclipse suggests apsidal motion, highlighting the system’s evolving orbital dynamics. (Fig.~\ref{lc:HD 278236}). HD 93683 exhibited a decrease in eclipse depth over 1500 days, suggesting changes in the system's configuration or surrounding material (Fig.~\ref{lc:HD_93683 unfolded}). And BD+66 1675 (detailed in \ref{subsec:NewDEB_discovered}, Fig.~\ref{lc:BD+66 1675_E}) presented a heartbeat-like feature post-primary eclipse, which shifted to prior the eclipse over 1150 days of observation.

\textbf{HD 93683}: 
It is recognized as a SB2 system with a Be-type third component \citep{2016AJ....152..190A,2020A&A...641A..42B}. Our analysis of the TESS data unveiled intriguing behavior in this system, characterized by an attenuation of its eclipses (Fig.~\ref{lc:HD_93683 unfolded}). This phenomenon could arise from variations in the brightness of the variable Be-star (either the star itself or its surrounding disk), which may dilute the eclipses. Alternatively, it could result from a rare effect known as Zeipel-Lidov-Kozai cycles, induced by the third component in a noncoplanar orbit, leading to changes in the orbital plane relative to our line of sight. As far as we know, this is the very first case reported in massive systems \citep[see e.g.,][and references therein]{2022MNRAS.515.3773B}.
This multiple system also exhibits several short-period variations, interpreted as pulsations. \citet{2022ApJS..259...50S} reported one such variation with a period of $2.4$ days. Further analysis is needed to determine additional periodicities in the TESS data.

\textbf{* eta Ori}: 
*Eta Ori is an ideal candidate for our catalog as it meets the criteria of a detached binary with spectral types B0.7 V and B1.5: V. \cite{2022MNRAS.513.3191S} analyzed TESS data and fitted the LC to obtain absolute parameters for the system, opting not to use data from sector 32 due to its low variability amplitude. In our study, we present LCs using both sectors (Fig.~\ref{lc:eta_OriJ}). We confirm the periodic variations reported in the literature: the primary eclipse cycle at approximately 7.989 days, indicative of the detached eclipsing binary (EB) system, and a shorter cycle of about 0.432 days, associated with pulsational variability. \cite{1993ASPC...38..239L} initially suggested that the shorter period might indicate a contact binary with a period of 0.864 days. Southworth proposed the configuration as a detached EB with a period of 7.988 days, where one component exhibits g-mode pulsations, alongside a noneclipsing binary with a period of 0.8641 days, showing strong ellipsoidal variations.

\section{The Young Massive Detached Binary catalog}
This work presents the comprehensive YMDB catalog, derived from the analysis of TESS LCs and spectroscopic data, with additional support from an extensive review of existing literature. 

Eclipsing binaries offer a unique opportunity to determine stellar parameters with high precision, especially when combining LC information with RV data. Detached binaries are particularly valuable due to the minimal interaction between their stellar components, allowing for accurate determinations of stellar parameters. 

Although systems within the O8–B3 spectral-type range are common, few have had their absolute parameters precisely measured. The YMDB catalog addresses this knowledge gap by providing a curated database of young massive detached binaries, facilitating high-precision stellar parameter determinations. 

Through the analysis of TESS LCs for 87 systems with suspected spectral types in the range O9-B1, this study identified 20 new eclipsing binaries, including 13 previously unknown variable systems and 2 nonthermal contact binaries. Additionally, new LC classifications were reported for 30 systems, and novel features such as eccentricity and heartbeat phenomena were discovered in many targets. 

The YMDB catalog offers a reliable resource of high-quality LCs, serving as a valuable asset for the astronomical community. The primary results of this study are documented in Table~\ref{tab:YMDB_confirmed}, which lists the 30 confirmed members of the YMDB. These systems feature detached LCs and have at least one component with a spectral classification within the specified range. For the 25 systems that show potential yet require further spectroscopic verification, details can be found in Table~\ref{tab:YMDB_candidates}. The 32 systems that do not qualify for the catalog due to nondetached configurations or incompatible spectral types are included in Table~\ref{tab:YMDB_unqualified}

It is our desire to further calibrate this method and publish a semiautomatic pipeline for public use.

\begin{table*}
    \centering
    \hypersetup{citecolor=black}
    \caption{Confirmed systems in the YMDB catalog.}
    \label{tab:YMDB_confirmed}
    \begin{adjustbox}{max width=\textwidth}
    \begin{tabular}{lcccc ccccc ccc}
    
\hline
SIMBAD         & $\Delta$Mag    & P                    & T0                      & Apsidal & Multi\_P & HB    & e     & ET             & ST1                         & ST2                       \\%
\hline
V* NY Cep      & 0.140         & 15.2759              & 1768.17910              &         & 0        & 1     & 1     & EA1            & B0.5V                       & B2V                       \\%
HD 99898       & 0.206         &  5.04950             & 2324.03500              & 1       & 0        & 1     & 1     & EA2            & B0V                         &                           \\%
V* V404 Vel    & 0.103         & 11.4243              & 1518.04660              &         & 0        & 0     & 1     & EA2            & B0V                         &                           \\%
HD 298448      & 0.142         &  2.31530             & 1545.16410              &         & 0        & 0     & 1     & EA2            & B1 V\textsuperscript{ tw}   &                           \\%
V* KU Car      & 0.574         &  2.96059             & 1571.58360              &         & 0        & 0     & 0     & EA2            & B0.5 V(n)                   &                           \\%
BD+66 1674     & 0.123/0.017   & 18.8292  / 2.6395    & 1783.81040 / 2887.35633 &         & 0 / 1    & 0 / 1 & - / 1 & EA1 / EA2      & B0 V\textsuperscript{ tw}   & B0 V\textsuperscript{ tw} \\%
V* V1208 Sco   & 0.124         &  5.21970             & 2387.49300              &         & 0        & 0     & 0     & EA2            & B0.5 V                      & *                         \\%
V* AH Cep      & 0.308         &  1.77478             & 1792.24979              &         & 0        & 0     & 0     & EA2            & B0.2 V                      & B2 V                      \\%
V* DW Car      & 0.647         &  1.32774             & 1597.83274              &         & 0        & 0     & 0     & EA2            & B1 V                        & B1 V                      \\%
* eta Ori      & 0.308 / 0.061 &  7.9886  / 0.4321    & 1472.43778 / 1469.94135 &         & 1        & 0     & 1 / 0 & EA2 / EW       & B0.7 V                      & B1.5: V                   \\%
* del Pic      & 0.243         &  1.67254             & 2389.11655              &         & 1:       & 0     & 0     & EA2            & B0.5 V                      & B0.5-3                    \\%
V* VV Ori      & 0.310         &  1.48538             & 2199.25573              &         & 1        & 0     & 0     & EA2            & B1 V                        & B4.5 V                    \\%
HD 338936      & 0.417         &  7.670               & 2771.85150              &         & 1        & 0     & 1     & EA2            & B0.5 V db                   &                           \\%
TYC 8174-540-1 & 0.162         &  5.04690             & 1546.33040              &         & 1        & 1     & 1     & EA2            & O9.5 V\textsuperscript{ tw} &                           \\%
LS VI +00 25   & 0.057         &  11.0287             & 2221.42860              &         & 0        & 0     & 1     & EA2            & O9.5 V                      &                           \\%
HD 278236      & 0.157         &  1.99270             & 1816.47898              & 1       & 0        & 0     & 1     & EA2            & O9 V                        &                           \\%
CD-28 5257     & 0.196         &  3.10894  / 3.38:    & 2253.54773 / 1498.46100 & 1       & 1        & 1     & 1     & EA2            & B0 V\textsuperscript{ tw}   &                           \\%
HD 93683       & 0.041:        & 17.7978              & 1574.31401              & 1       & 1        & 0     & 1     & EA2            & O9 V                        & B0 V                      \\%
HD 152218      & 0.089         &  5.60410             & 1649.95040              &         & 1        & 1     & 1     & EA1            & O9 IV                       & O9.7 V                    \\%
V* V346 Cen    & 0.293         &  6.322               & 1570.60040              &         & 1        & 1     & 1     & EA2            & B0.5 IV                     & B2 V                      \\%
CD-35 4470     & 0.372         &  9.3535              & 1539.49440              &         & 0        & 0     & 1     & EA2            & B0 IV                       &                           \\%
HD 309036      & 0.232         &  2.31525 / ~0.12398  & 2356.19344              &         & 1        & 0     & 0     & EA2            & B1 V\textsuperscript{ tw}   &                           \\%
V* IK Vel      & 0.874         &  1.99232             & 1518.81743              &         & 0        & 0     & 0     & EA2            & B1 V\textsuperscript{ tw}   &                           \\%
RAFGL 5223     & 0.053         &  3.55907             & 1494.99690              & 1:      & 0        & 1     & 1     & EA2            & O9 V\textsuperscript{ tw}   & *                         \\%
Schulte 27     & 0.629         &  1.46920             & 1683.74890              &         & 0        & 0     & 0     & EA2            & O9.7V(n)                    & O9.7V(n)                  \\%
V* HH Car      & 0.281         &  3.23146             & 1599.54000              &         & 0        & 0     & 0     & EA2            & O9 V                        & B0 III-IV                 \\%
V* V1295 Sco   & 0.218         &  2.15764             & 1627.79500              &         & 0        & 0     & 0     & EA2            & O9.7 V                      & *                         \\%
V* V725 Car    & 0.152         &  9.4106              & 1570.04690              &         & 0        & 1     & 1     & EA2            & O9.7 IV                     & *                         \\%
V* Y Cyg       & 0.614         &  2.99625             & 1711.75278              & 1       & 0        & 1     & 1     & EA2            & O9.5 IV                     & O9.5 IV                   \\%
HD 204827      & 0.011         &  3.0480              & 1743.64949              &         & 1        & 0     & 0     & EA2            & O9.5IV                      & *                         \\%
\hline
    
    \end{tabular}
\end{adjustbox}
    \begin{justify}
        \textbf{Note:} Summarized version of the YMDB catalog for confirmed systems. The table includes Delta magnitude ($\Delta$mag), orbital Period (P), and Time of minimum light (T0) from extracted TESS LCs. Identified features such as Apsidal motion, additional variability (Multi\_P), Heartbeat-like features (HB), and Eccentricity (e) are presented with ``1'' indicating detection and ``0'' indicating nondetection. The Eclipsing Type (ET) is indicated as EA (Algol), EB (Beta Lyrae), or EW (W Ursa Majoris type), followed by ``2' if both primary and secondary eclipses are visible, or ``1'' if only one is visible. A ``/'' between types indicates multiple discernible variations from the TESS LC. Spectral types for primary and secondary components are provided, with classifications performed in this work marked with ``\textsuperscript{tw}''. An asterisk ``*'' instead of a spectral type indicates significant dispersion among various reliable sources, without a clear consensus. Uncertainties across the table are indicated by ``:''. Full details, error margins, and sources for spectral types not classified in this work are available in the CDS extended version.
    \end{justify}
\end{table*}

\begin{table*}
    \centering
    \hypersetup{citecolor=black}
    \caption{Unconfirmed systems (candidates) for the YMDB catalog.}
    \label{tab:YMDB_candidates}
    \begin{adjustbox}{max width=\textwidth}
    \begin{tabular}{lcccc ccccc ccc}

\hline
SIMBAD                  & $\Delta$Mag      & P                  & T0                      & Apsidal & Multi\_P & HB    & e     & ET        & ST1                          & ST2                        \\%
\hline
2MASS J16542949-4139149 & 0.182          &  6.34877           & 1633.56460              &         & 1        & 0     & 0     & EA2       & *                            &                            \\%
HD 114026               & 0.230          &  2.1620            & 2334.09192              &         & 0        & 0     & 0     & EA2       & B0.5 V:n                     &                            \\%
CPD-58 2608A            & 0.052          &  2.23284           & 2331.83340              & 1:      & 1:       & 0     & 1     & EA2       & *                            & *                          \\%
V* V1153 Cen            & 0.557          &  5.979             & 2335.72810              &         & 0        & 0     & 1     & EA2       & *                            &                            \\%
V* V1765 Cyg            & 0.164          & 13.3724            & 2794.57244              &         & 1        & 1     & 1     & EA2       & B0.5Ib                       & B2:V:                      \\%
CD-27 4726              & 0.120          &  7.9587            & 1498.98070              &         & 0        & 1:    & 1:    & EA1       & *                            &                            \\%
HD 306096               & 0.304          &  5.38300           & 1574.47920              &         & 0        & 1     & 1     & EA2       & B0                           &                            \\%
V* GN Nor               & 0.559          &  5.703             & 1626.08130              &         & 0        & 1     & 1     & EA2       & A0 V:\textsuperscript{ tw}   &                            \\%
V* V1103 Cas            & 0.566          &  6.178             & 1790.53260              &         & 0        & 0     & 1     & EA2       & B0                           &                            \\%
CD-59 3165              & 0.240 / 0.015: &  7.59370 / 3.18206 & 1573.18250 / 1616.16205 &         & 0 / 0    & 0 / 0 & 1 / 0 & EA2 / EA2 & *                            &                            \\%
V* CE CMa               & 0.638          & 27.0729            & 2247.82650              &         & 0        & 0     & 1     & EA2       & *                            &                            \\%
HD 52504                & 0.381          &  1.42147           & 2225.87947              &         & 0        & 0     & 0     & EA2       & B1: V:                       & -                          \\%
V* V646 Cas             & 0.471          &  6.16200           & 2006.30329              &         & 0        & 0     & 0     & EA2       & B0 IV:nn                     &                            \\%
CPD-42 2880             & 0.095          &  1.8988            & 1566.58619              &         & 0        & 0     & 0     & EA2       & O9.5-B2                      &                            \\%
CPD-63 3284             & 0.167          &  2.872             & 2359.61038              &         & 1        & 0     & 0     & EA2       & OB                           &                            \\%
HD 111825               & 0.227          &  2.00669           & 1596.80501              &         & 0        & 0     & 0     & EA2       & *                            &                            \\%
CD-54 6456              & 0.097          & 16.9742            & 1634.97340              &         & 0        & 0     & 1     & EA2       & *                            & *                          \\%
HD 102475               & 0.012          &  9.0411            & 2324.78323              &         & 1        & 0     & -     & EA1       & *                            & *                          \\%
HD 152219               & 0.192          &  4.24024           & 2364.88748              &         & 1        & 1     & 1     & EA2       & O9.5 III                     & B1-2 V-III                 \\%
V* EV Vul               & 0.863          &  2.82212           & 2421.09141              &         & 1        & 0     & 0     & EA2       & *                            & *                          \\%
V* V499 Sco             & 0.501          &  2.33329           & 2389.56663              &         & 0        & 0     & 0     & EA2       & *                            & *                          \\%
* f Vel                 & 0.086          & 26.3060            & 2302.27932              &         & 0        & 1     & 1     & EA1       & O9.7 II\textsuperscript{ tw} & B0 V:\textsuperscript{ tw} \\%
V* XZ Cep               & 0.863          &  5.0973            & 1982.06212              &         & 0        & 0     & 0     & EA2       & B1.5 II-III                  & B1.1 III-V                 \\%
BD+66 1675              & 0.029          &  2.71564           & 1768.18230              &         & 1        & 1:    & 1:    & EA1       & O8 V                         & B                          \\%
BD+55 2722              & 0.036          &  2.00453           & 2881.12965              &         & 1        & 0     & 0     & EA2       & O7 Vz(n)                     & B                          \\%
\hline

    \end{tabular}
\end{adjustbox}
    \begin{justify}
        \textbf{Note:} Summarized version of the YMDB catalog for unconfirmed systems (candidates). Presented in the same format as Table~\ref{tab:YMDB_confirmed}.
    \end{justify}
\end{table*}

\begin{table*}
    \centering
    \hypersetup{citecolor=black}    
    \caption{Unqualified systems for the YMDB catalog.}
    \label{tab:YMDB_unqualified}
    \begin{adjustbox}{max width=\textwidth}
    \begin{tabular}{lcccc ccccc ccc}
    
\hline
SIMBAD        & $\Delta$Mag & P                       & T0         & Apsidal & Multi\_P & HB & e   & ET   & ST1                           & ST2                        \\%
\hline
V* AC Vel     & 0.442      &  4.562                  & 2284.25847 &         & 0        & 0  & 0   & EA2  & B2 V\textsuperscript{ tw}     &                            \\%
HD 52533      & 0.447      & 21.9648                 & 1501.65900 &         & 0        & 0  & 1   & EA2  & O8.5IVn                       &                            \\%
* 23 Ori      & 0.039      &  4.55520                & 1470.80658 &         & 1        & 0  & 0   & EA2  & B2 IV/V                       & *                          \\%
HD 309018     & 0.052:     &  0.89216                & 2358.72842 &         & 0        & 0  & 1   & EA2  & O8.5 V\textsuperscript{ tw}   &                            \\%
HD 144918     & 0.059      &  1.27913                & 1625.75646 &         & 0        & 0  & 0   & EA2  & O8 V\textsuperscript{ tw}     & *                          \\%
V* V340 Mus   & 0.076      &  3.42725                & 1597.38243 &         & 1        & 0  & 0   & EA2  & O9 IV                         & *                          \\%
V* FM CMa     & 0.239      &  2.78940                & 1509.69900 & 1       & 0        & 0  & 0   & EA2  & B2 IV\textsuperscript{ tw}    & B3 V:\textsuperscript{ tw} \\%
V* ET Vel     & 0.657      &  3.08090                & 1519.76470 &         & 0        & 0  & 1   & EA2  & B2.5 V\textsuperscript{ tw}   &                            \\%
HD 305850     & 0.086      &  2.3810                 & 1620.65763 &         & 0        & 0  & 0   & EB2  & *                             &                            \\%
V* V1082 Sco  & 0.402      & 23.4465                 & 1637.81050 &         & 1        & 0  & 1   & EA2  & B0.5Ib                        & O9.5 III                   \\%
HD 99630      & 0.214      &  21.1220                & 1602.84699 &         & 0        & 0  & 1   & EA2  & B4-B5                         &                            \\%
UCAC2 5911156 & 0.189      &  8.6715                 & 2379.81580 &         & 1        & 0  & -   & EA1  & B0.5 III\textsuperscript{ tw} &                            \\%
CD-53 6352    & 0.034      &  2.48610                & 1630.08061 &         & 1        & 0  & 0   & EA2  & O7III (for component A)       &                            \\%
CPD-64 1885   & 0.292      &  5.13930                & 1600.42950 &         & 0        & 0  & 0   & EA2  & B4-B6                         &                            \\%
HD 277878     & 0.019      &  0.82253                & 2934.44773 &         & 1:       & 0  & 0   & EA2  & O7 V((f))z                    &                            \\%
CD-59 5583    & 0.149      &  8.7060                 & 2387.67312 &         & 1        & 1  & 1   & EA2  & B0II                          &                            \\%
HD 338961     & 0.186      &  1.70970                & 2793.63189 &         & 0        & 0  & 0   & EA2  & B0.5 Illnn                    &                            \\%
V* V1216 Sco  & 0.499      &  3.92060                & 2389.62316 &         & 1        & 0  & 0   & EA:2 & B0.5III                       &                            \\%
V* V421 Pup   & 0.137      &  5.41650                & 1500.20265 &         & 1        & 0  & 0   & EA2  & B1II                          &                            \\%
HD 103223     & 0.332      &  2.54155                & 2358.08085 &         & 0        & 0  & 0   & EA2  & B2.5 V\textsuperscript{ tw}   &                            \\%
V* V1290 Sco  & 0.116      &  4.49260                & 1627.90670 &         & 1        & 0  & 0   & EA2  & O9.7 III                      & *                          \\%
V* V4386 Sgr  & 0.231      & 10.802                  & 1664.15850 &         & 1        & 0  & 0   & EA2  & B0.5 III\textsuperscript{ tw} & *                          \\%
HD 142152     & 0.139      &  5.68640                & 1626.34840 & 1       & 1        & 1  & 1   & EA2  & B0 III\textsuperscript{ tw}   & *                          \\%
HD 37737      & 0.119      &  7.85199  / 1/10P / ~3P & 1821.42367 &         & 1        & 1  & 1   & EA2  & O9.5II-III(n)                 & *                          \\%
V* V399 Pup   & 0.204      &  3.91019                & 2276.30939 & 1       & 0        & 1  & 1   & EA2  & B2 II                         & *                          \\%
V* CC Cas     & 0.145      &  3.3670                 & 1818.57209 &         & 1        & 0  & 0   & EA2  & O8.5III(n)((f))               & *                          \\%
V* MN Cen     & 0.587      &  3.48915                & 2359.82698 &         & 0        & 0  & 0   & EA2  & B1.5 V\textsuperscript{ tw}   & *                          \\%
V* V877 Cen   & 0.636      &  5.35857                & 2355.76939 &         & 0        & 0  & 0   & EA2  & B1III(n)                      & *                          \\%
* psi02 Ori   & 0.037      &  2.52596                & 2199.90592 &         & 1        & 0  & 1   & EA2  & B1 III                        & B2 V                       \\%
* u Her       & 0.663      &  2.05102                & 2767.92462 &         & 0        & 0  & 0   & EA2  & B2 IV                         & B8 III                     \\%
LS V +38 12   & 0.053      &  1.42287                & 1837.24509 &         & 0        & 0  & 0   & EB2  & O7 V ((f))                    & B0III- V                   \\%
HD 185780     & 0.024      &  3.51160                & 2822.28161 &         & 1        & 0  & 0   & EA2  & B0 III                        & *                          \\%
\hline

    \end{tabular} \par
\end{adjustbox}
    \begin{justify}
        \textbf{Note:} Summarized version of the YMDB catalog for unqualified systems. Presented in the same format as Table~\ref{tab:YMDB_confirmed}.
    \end{justify}    
    
\end{table*}

\begin{acknowledgements}
RG acknowledges support from grant PICT 2019-0344. JIA acknowledges the financial support of DIDULS/ULS, through the project PR2324063. This research made use of Lightkurve, a Python package for Kepler and TESS data analysis (Lightkurve Collaboration, 2018). This work made use of Astropy:\footnote{http://www.astropy.org} a community-developed core Python package and an ecosystem of tools and resources for astronomy \citep{2022ApJ...935..167A}.
\end{acknowledgements}

\bibliographystyle{aa}


\begin{appendix}
\section{Folded light curves for all 87 systems analyzed in this study}
\begin{figure*}
	\includegraphics[width=2\columnwidth]{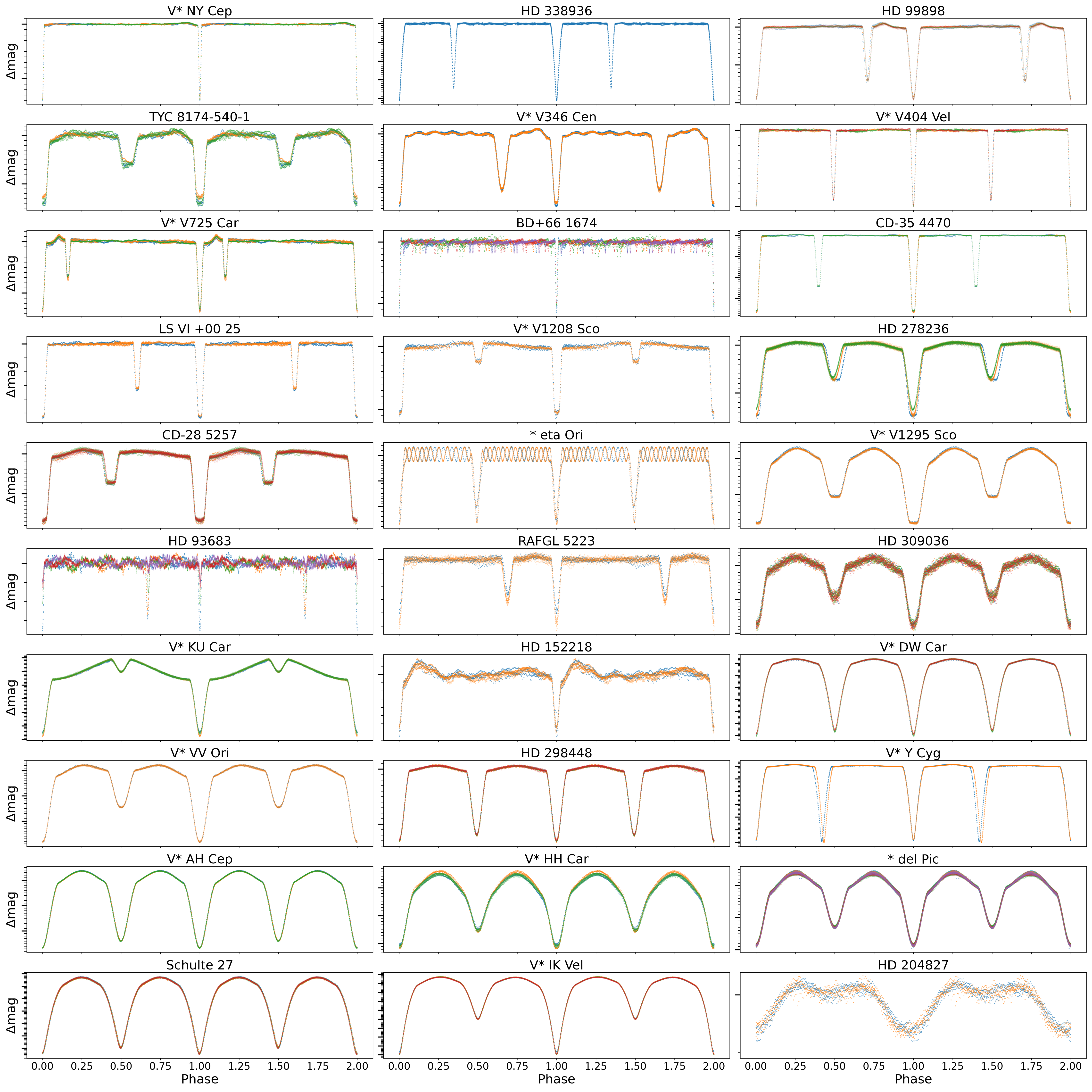} 
	\caption{Light curves belonging to the confirmed category. In each subpanel, bold, long ticks on the Y-axis denote increments of 0.1$\Delta mag$, and thin, short ticks indicate increments of 0.01$\Delta mag$.}
	\label{lc:confirmedAll}
\end{figure*}

\begin{figure*}
	\includegraphics[width=2\columnwidth]{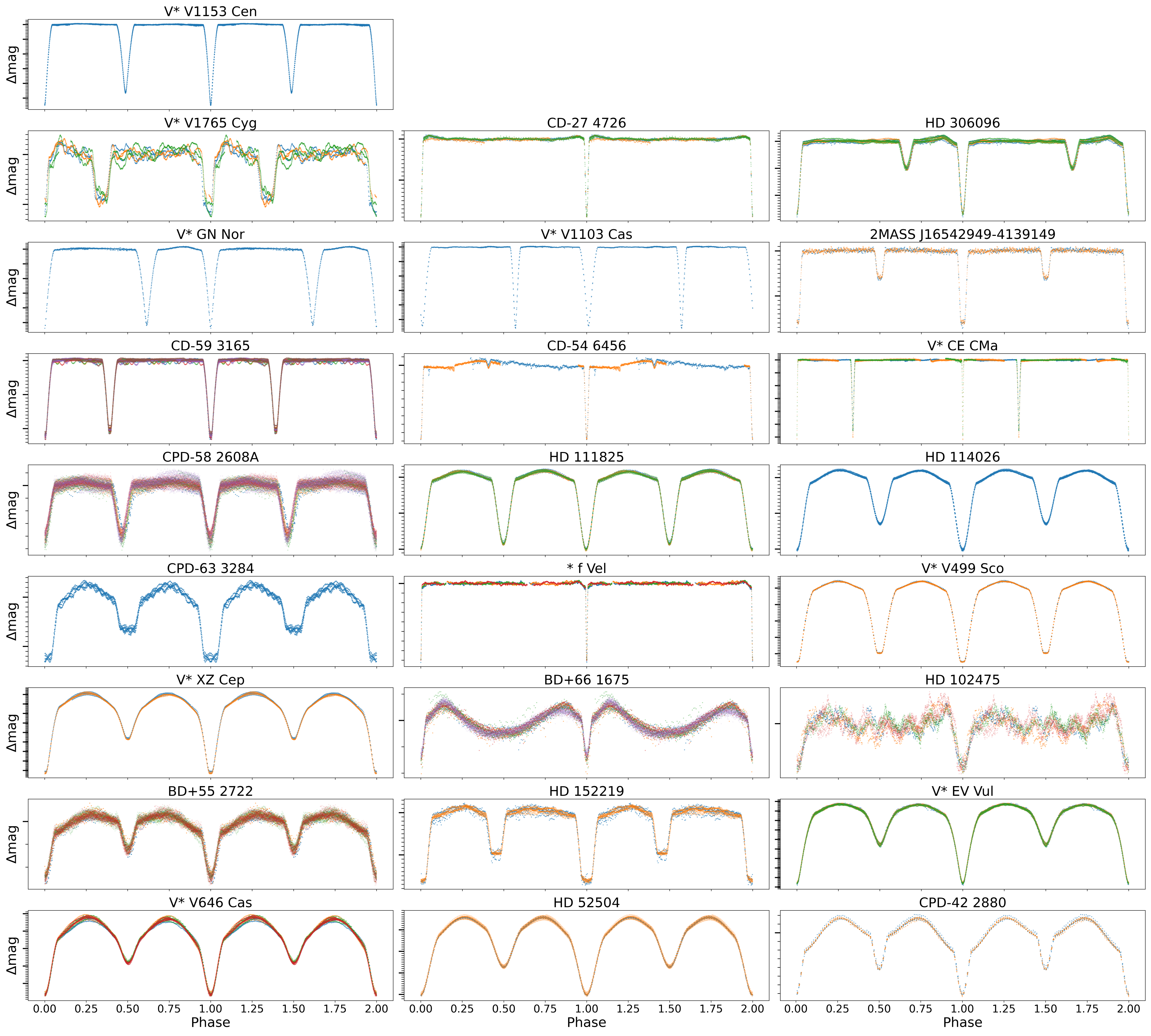} 
	\caption{Light curves belonging to the candidate category. In each subpanel, bold, long ticks on the Y-axis denote increments of 0.1$\Delta mag$, and thin, short ticks indicate increments of 0.01$\Delta mag$.}
	\label{lc:candidatesAll}
\end{figure*}

\begin{figure*}
	\includegraphics[width=2\columnwidth]{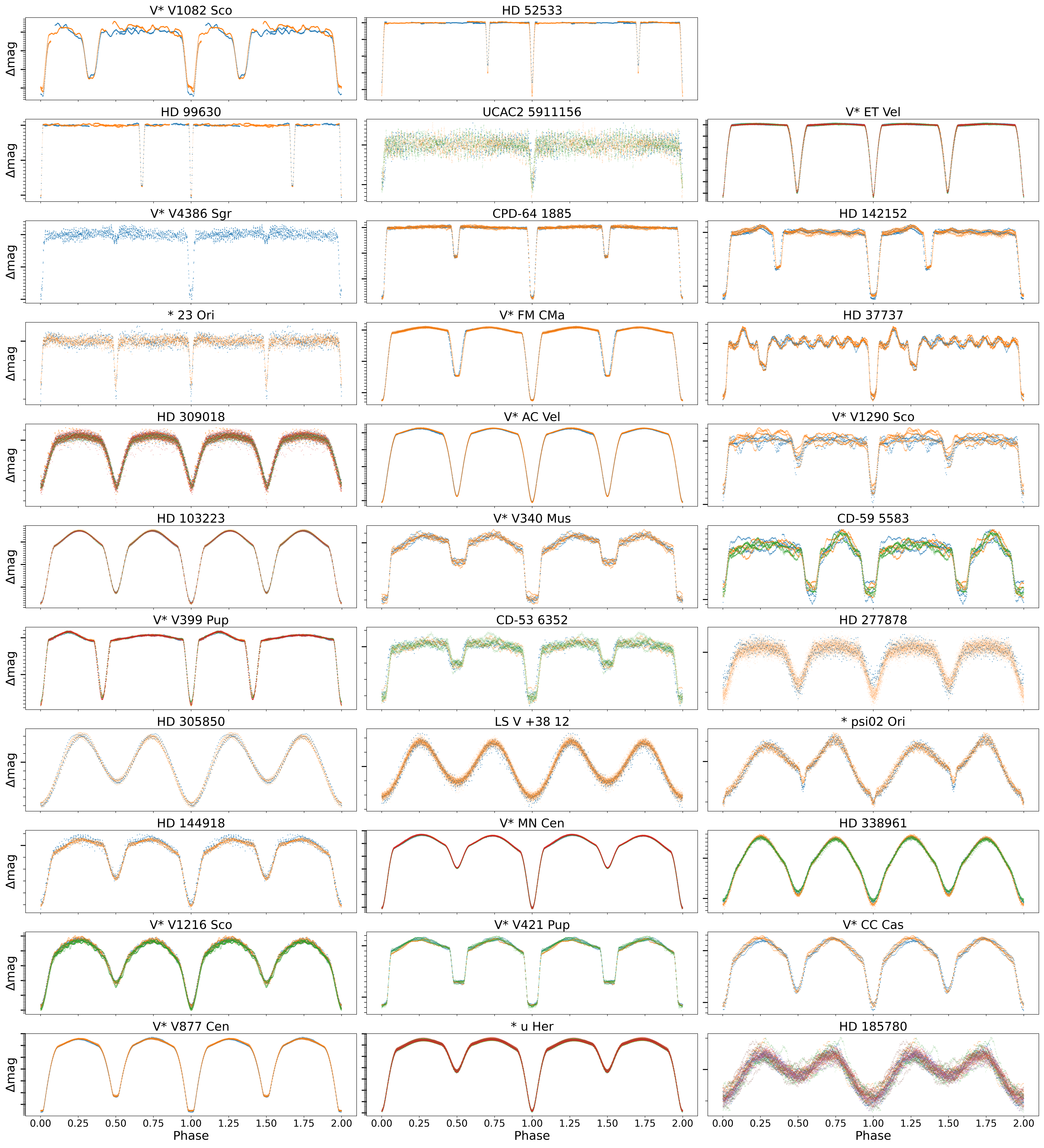} 
	\caption{Light curves belonging to the unqualified category. In each subpanel, bold, long ticks on the Y-axis denote increments of 0.1$\Delta mag$, and thin, short ticks indicate increments of 0.01$\Delta mag$.}
	\label{lc:unqualifiedAll}
\end{figure*}

The appendix presents the LCs for all systems analyzed in this study. Figure \ref{lc:confirmedAll} shows the LCs for the confirmed systems, Figure \ref{lc:candidatesAll} displays the LCs for the candidate systems, and Figure \ref{lc:unqualifiedAll} includes the LCs for the unqualified systems.

\end{appendix}

\end{document}